\documentclass[useAMS,usenatbib]{mn2e}
\usepackage{graphics}
\usepackage{graphicx}
\usepackage{psfig}
\input{epsf}
\def\half{{1\over 2}}

\def\etal{{\it et~al.\ }}

\def\ie{{\it i.e.~}}

\def\ltsima{$\; \buildrel < \over \sim \;$}
\def\simlt{\lower.5ex\hbox{\ltsima}}
\def\gtsima{$\; \buildrel > \over \sim \;$}
\def\simgt{\lower.5ex\hbox{\gtsima}}
\def\statei{\lbrace r,s \rbrace}
\def\statej {\lbrace r,q \rbrace}
\def\statek{\lbrace s,q \rbrace}

\def \lleq {\lower0.9ex\hbox{ $\buildrel < \over \sim$} ~}
\def \ggeq {\lower0.9ex\hbox{ $\buildrel > \over \sim$} ~}
\def\atridot{\stackrel{...}{a}}
\def\l{\Lambda}
\def\o{\Omega}
\def\omx {\Omega_X}
\def\om {\Omega_m}

\def \rb{\bar{r}}
\def \sb{\bar{s}}
\def \qb{\bar{q}}
\def\beq{\begin{equation}}
\def\eeq{\end{equation}}

\def\ber{\begin{eqnarray}}
\def\eer{\end{eqnarray}}

\def\apj{{Astroph.\@ J.\ }}
\def\mn{{Mon.\@ Not.\@ Roy.\@ Ast.\@ Soc.\ }}
\def\asta{{Astron.\@ Astrophys.\ }}
\def\aj{{Astron.\@ J.\ }}
\def\prl{{Phys.\@ Rev.\@ Lett.\ }}
\def\pd{{Phys.\@ Rev.\@ D\ }}

\def\plb {{Phys.\@ Lett.\@ B\ }}
\def \jetpl {JETP Lett.\ }

\begin{document}

\title{Exploring the Expanding Universe and Dark Energy using 
the Statefinder Diagnostic}

\author[Ujjaini Alam, Varun Sahni, Tarun Deep Saini, A. A. Starobinsky]{Ujjaini Alam$^{1,4}$, Varun Sahni$^{1,5}$, Tarun Deep Saini$^{2,6}$ and A. A. Starobinsky$^{3,7}$ \\
  $^{1}$ Inter University Centre for Astronomy $\&$ Astrophysics, Pune, India\\
  $^{2}$ Institute of Astronomy, Madingley Road, Cambridge, UK \\
  $^{3}$ Landau Institute for Theoretical Physics, 119334 Moscow, Russia\\
  $^{4}$ujjaini@iucaa.ernet.in \\
  $^{5}$varun@iucaa.ernet.in \\
  $^{6}$tarun@ast.cam.ac.uk \\
  $^{7}$alstar@landau.ac.ru \\
  } \maketitle


\begin{abstract}
The coming few years are likely to witness a dramatic increase in high
quality Sn data as current surveys add more high redshift supernovae
to their inventory and as newer and deeper supernova experiments
become operational.  Given the current variety in dark energy models
and the expected improvement in observational data, an accurate and
versatile diagnostic of dark energy is the need of the hour.  This
paper examines the Statefinder diagnostic in the light of the proposed
SNAP satellite which is expected to observe about 2000 supernovae per
year.  We show that the Statefinder is versatile enough to
differentiate between dark energy models as varied as the cosmological
constant on the one hand, and quintessence, the Chaplygin gas and
braneworld models, on the other.  Using SNAP data, the Statefinder can
distinguish a cosmological constant ($w=-1$) from quintessence models
with $w \geq -0.9$ and Chaplygin gas models with $\kappa \leq 15$ at
the $3\sigma$ level if the value of $\om$ is known exactly. The
Statefinder gives reasonable results even when the value of $\om$ is
known to only $\sim 20\%$ accuracy.  In this case, marginalizing over
$\om$ and assuming a fiducial LCDM model allows us to rule out
quintessence with $w \geq -0.85$ and the Chaplygin gas with $\kappa
\leq 7$ (both at $3\sigma$).  These constraints can be made even
tighter if we use the Statefinders in conjunction with the
deceleration parameter. The Statefinder is very sensitive to the
total pressure exerted by all forms of matter and radiation in the
universe. It can therefore differentiate between dark energy models at
moderately high redshifts of $z \lleq 10$.
\end{abstract}

\section{Introduction }

Supernova observations \citep{riess,perl}, when combined with those of
the cosmic microwave background \citep{archeops} and gravitational
clustering \citep{lss}, suggest that our Universe is (approximately)
spatially flat and that an exotic form of negative-pressure matter
called `dark energy' (DE) causes it to accelerate by contributing as
much as 2/3 to the closure density of the universe -- the remaining
third consisting of non-relativistic dark matter and baryons.  The
simplest example of dark energy is the cosmological constant ($\l$),
with associated mass density
\beq
\rho_{\l}=6.44 \times 10^{-30}\left({\Omega_{\l}\over 0.7}\right)
\left({h\over 0.7}\right)^2 {\rm {g~cm^{-3}}}\,\,,
\label{Lambda}
\eeq 
where $h$ is the Hubble constant $H_0$ in terms of $100$ km s$^{-1}$
Mpc$^{-1}$ and $\Omega_{\l}=0.7\pm 0.1,~h= 0.7\pm 0.1$.  Although the
cold dark matter model with a cosmological constant (hereafter LCDM)
provides an excellent explanation for the acceleration phenomenon and
other existing observational data, it remains entirely plausible that
the dark energy density is weakly time dependent \citep[see the
reviews][]{ss00,pr02}. Moreover, it is natural to suggest (in complete
analogy with what has been done in the case of another type of `dark
energy' responsible for driving the expansion of the universe during
an inflationary stage in the early universe) that the dark energy
which we observe today might really be dynamical in nature and origin.
This means that a completely new form of matter is responsible for
giving rise to the second inflationary regime which we are entering
now.

Many models of dark energy have been proposed; in fact, any
inflationary model (even a `bad' one, i.e. without a `graceful exit'
to the subsequent radiation-dominated Friedmann-Robertson-Walker (FRW)
stage) may be used for this purpose if one assumes different values
for its microscopic parameters. The simplest of these models rely on a
scalar field minimally interacting with Einstein gravity --
quintessence \citep{ratra88,peebles88,frieman95,cald98}, and bear an
obvious similarity with the simplest variants of the inflationary
scenario. Inclusion of a non-minimal coupling to gravity in these
models together with further generalization leads to models of dark
energy in a scalar-tensor theory of gravity \citep[see][and references
therein]{beps00}.  Other models invoke matter with unusual properties
such as the Chaplygin gas \citep{chap1} or k-essence \citep{mukhanov}.
Still others generate cosmic acceleration through topological defects
\citep{spergel98} or quantum vacuum polarization and particle
production \citep{sahni98,parker99}.  Lately it has been noticed that
higher dimensional `braneworld' models could account for a late-time
accelerating phase even in the absence of matter violating the strong
energy condition \citep{dgp00,DDG,DDG1,ss02a,as02} \citep[see][for a
recent review of dark energy models]{sahni02}. It is especially
interesting that in the latter class of models 'dark energy' need not
be an energy of some form of matter at all, but can have an entirely
geometrical origin. Moreover, in these models the basic gravitational
field equations do not have the Einstein form
\beq
R_{\alpha \beta}-{\half}g_{\alpha \beta}R=8 \pi G \left( T_{\alpha
    \beta} \big \vert_{\mathit matter}+T_{\alpha \beta} \big \vert_{\mathit
    radiation}+T_{\alpha \beta} \big \vert_{\mathit DE}\right)
\label{Eform}
\eeq 
($c=\hbar =1$ is assumed here and below), and therefore the notions of
`energy density' and `pressure' of DE loose their exact fundamental
sense and become ambiguous and convention-dependent. A major ambiguity
arises in models of scalar-tensor gravity as well as in braneworld
models both of which contain interaction terms between dark energy and
non-relativistic matter. Interpreting such models within the Einstein
framework (\ref{Eform}) leads to the following dilemma: should these
interaction terms be ascribed to dark matter (hence to
$T_{\alpha\beta}\vert_{\mathit matter}$ in (\ref{Eform})) or to dark
energy (to $T_{\alpha\beta}\vert_{\mathit DE}$) ?  Our answer to this
question has the potential to alter the properties of dark energy
including its density and pressure and hence also its equation of
state.  In marked contrast to such ambiguities which could arise if we
are not careful with our usage of the term `equation of state', the
expansion factor of the universe in the physical frame $a(t)$, when
expressed through the Hubble parameter $H\equiv {\dot a\over a}$, is
an unambiguous, fundamental and readily measurable quantity.

Given the rapidly improving quality of observational data and also the
abundance of different theoretical models of dark energy, the need of
the hour clearly is a robust and sensitive statistic which can succeed
in differentiating cosmological models with various kinds of dark
energy both from each other and, even more importantly, from an exact
cosmological constant. In view of the non-fundamentality of the
notions of DE density and pressure pointed out above, we prefer to
work with purely geometric quantities. Then such a sensitive
diagnostic of the present acceleration epoch and of dark energy could
be the statefinder pair $\statei$, recently introduced in
\citet{sssa02}. The statefinder probes the expansion dynamics of the
universe through higher derivatives of the expansion factor ${\ddot a}
~\& \atridot$. Its important property is that $\statei=\lbrace 1,0
\rbrace$ is a fixed point for the flat LCDM FRW cosmological model.
Departure of a given DE model from this fixed point is a good way of
establishing the `distance' of this model from flat LCDM.  As we will
show in this paper, the statefinder successfully differentiates
between rival DE models and, when combined with SNAP supernova data,
can serve as a versatile and powerful diagnostic of dark energy.
  
The paper is organized as follows. In the next section we briefly
review some theoretical models of dark energy. The behaviour of the
statefinder pair for these models is discussed in Section III while
the nature of data expected to become available from the SNAP
experiment is the subject of Section IV. Section IV also discusses
model-independent parametric reconstructions of dark energy.  Our
conclusions are presented in section V.

\section{Dark energy models and the acceleration of the universe}
\label{sec:darkenergy}

The rate of expansion  of a FRW universe and its acceleration are
described by the pair of equations
\ber\label{eq:acc} 
H^2 &=& \frac{8\pi G}{3}\sum_i \rho_i - \frac{k}{a^2}~,\nonumber\\
\frac{\ddot a}{a} &=& -\frac{4\pi G}{3}\sum_i (\rho_i
+ 3p_i), 
\eer 
where the summation is over all matter fields contributing to the
dynamics of the universe.  Clearly a necessary (but not sufficient)
condition for acceleration (${\ddot a} > 0$) is that at least one of
the matter fields in (\ref{eq:acc}) violate the strong energy
condition $\rho+3p \geq 0$.  If for simplicity we assume that the dark
energy pressure and density are related by the simple linear relation
$p = w\rho$, then $w < -1/3$ is a necessary condition for the universe
to accelerate. The acceleration of the universe can be quantified
through a dimensionless cosmological function known as the
`deceleration parameter' $q = - {\ddot a}/aH^2$, equivalently
\beq \label{eq:decel}
q(x) = \frac{H'(x)}{H(x)}~ x - 1 \, , ~~~ x = 1 + z \, ,
\eeq
where $q < 0$ describes an accelerating universe, whereas $q \geq 0$
for a universe which is either decelerating or expanding at the
`coasting' rate $a \propto t$.  As it will soon be shown, the
deceleration parameter on its own does not characterize the current
accelerating phase uniquely.  The presence of a fairly large
degeneracy in $q(z)$ is reflected in the fact that rival dark energy
models can give rise to one and the same value of $q_0$ at the present
time.  This degeneracy is easily broken if, as demonstrated in section
\ref{sec:statefinder}, one combines $q(z)$ with one of the
statefinders $r(z)$, $s(z)$.  The diagnostic pairs $\statej$ and
$\statek$ provide a very comprehensive description of the dynamics of
the universe and consequently of the nature of dark energy.

Now let us come to the issue of defining the energy density and
pressure of DE. In view of the ambiguities discussed in the
Introduction, we shall define $\rho_{\rm DE}$ and $p_{\rm DE}$ by
making use of the {\it Einstein interpretation} of gravitational field
equations (not to be confused with the notion of the Einstein frame
which is used in scalar-tensor and string theories of gravity!).
Namely, we assume that the gravitational field equations in a
single-metric theory of 3+1 gravity can be formally written in the
form (\ref{Eform}) where the Einstein tensor standing in the left-hand
side is defined with respect to the physical space-time metric. {\em
  All other terms are transferred to the right-hand side.}  Next, we
subtract the energy-momentum tensor of dust (CDM + baryons) from the
total energy-momentum tensor of matter and call the remaining part
`the effective energy-momentum tensor of dark energy' (in the Einstein
interpretation). Combining this prescription with Eq. (\ref{eq:acc})
and in the absence of spatial curvature, the energy density and
pressure of dark energy can be defined as:
\ber\label{eq:energy}
\rho_{\rm DE} &=& \rho_{\rm critical} - \rho_{\rm m} =
\frac{3H^2}{8\pi G}(1 - \Omega_{\rm m})~,\nonumber\\
p_{\rm DE} &=& \frac{H^2}{4\pi G}(q - \frac{1}{2})~,
\eer 
where $\rho_{\rm critical} = 3H^2/8\pi G$ is the critical density
associated with a FRW universe.  An important consequence of using
this approach is that the ratio $w_{DE}\equiv p_{DE}/\rho_{DE}$ can be
expressed in terms of the deceleration parameter
\beq\label{eq:state}
w_{\rm eff}(x) = {2 q(x) - 1 \over 3 \left( 1 - \Omega_{\rm m}(x) \right)}
\equiv \frac{(2 x /3) \ d \ {\rm ln}H \ / \ dx - 1}{1 \ - \ (H_0/H)^2 
\om \ x^3}\,\,.
\eeq
Following the above prescription we get standard results for the
cosmological constant and quintessence (for instance we recover Eq.
(\ref{eq:scalar})).  However the same cannot be said of braneworld
models since the Hubble parameter for the latter contains interaction
terms between matter and dark energy (see for instance Eqs.
(\ref{eq:ddg}), (\ref{eq:hubble_brane})) and therefore does not
subscribe to the Einsteinian format (\ref{Eform}) \& (\ref{eq:acc}).
One can however extend the above definition of $w_{\rm eff}$ to
non-Einsteinian theories by {\em defining} dark energy density to be
the remainder term after one subtracts the matter density from the
critical density in the Einstein equations.  It should be emphasised
that, according to this prescription all interaction terms between
matter and dark energy (such terms arise in scalar-tensor and brane
models) are attributed {\em solely} to dark energy. Therefore $w_{\rm
  eff}(z)$ defined according to (\ref{eq:state}) is an {\em effective
  equation of state} in these models and not a fundamental physical
entity (as it is in LCDM, for instance).

In this connection we should also stress that the propagation velocity
of small inhomogeneities in dark energy is generically neither
$\sqrt{w_{DE}}$, nor $\sqrt{dp_{DE} / d\rho_{DE}}$. Therefore although
$w(z)$ is an important physical quantity it does not provide us with
an exhaustive description of dark energy and its use as a diagnostic
should be treated with some caution.  (In this paper we will restrict
ourselves to a spatially flat FRW model and will not consider
inhomogeneous perturbations on this background.)

We now highlight a few popular candidates for dark energy which shall
be the focus of our discussion in this paper.

\begin{itemize}
  
\item {\bf Cosmological Constant.} Perhaps the simplest model for dark
energy is a cosmological constant $\Lambda$, whose energy density
remains constant with time $\rho_\l \equiv \l/8\pi G = -p_\l$, and
which has an equation of state $w_{\Lambda} =-1$.  A universe
consisting of matter in the form of dust and the cosmological
constant is popularly known as LCDM, the Hubble parameter for this
model has the form
\beq\label{eq:lambda}
H(z) = H_0 \left\lbrack \o_{\rm m}(1+z)^3 + 1 - 
\o_{\rm m}\right\rbrack^{1/2}~.
\eeq

\item{\bf Quiessence.} The next simplest form of dark energy after the
cosmological constant is provided by models for which the equation
of state is a constant $w = {\rm constant} \neq -1$.  For this form
of dark energy, which we call `quiessence' 
\beq\label{eq:quiessence}
H(z) = H_0 \left\lbrack \o_{\rm m}(1+z)^3 +
\omx(1+z)^{3(1+w)}\right\rbrack^{1/2}~.
\eeq
For $w=-1$ we recover the limiting form (\ref{eq:lambda}).  Important
examples of quiessence include a network of non-interacting cosmic
strings ($w = -1/3$) and domain walls ($w = -2/3$). Quiessence in a
FRW universe can also be produced by a scalar field (quintessence)
which has the potential $V(\phi)\propto \sinh ^{{-2(1+w)\over
    w}}(C\phi+D)$, with appropriately chosen values of $C$ and $D$
\citep[see][]{ss00,um00}.

Usually the dark energy equation of state depends upon time. We call 
such more generic models {\em kinessence}.

\item {\bf Quintessence} The simplest example of kinessence is
provided by quintessence -- a self-interacting scalar field which
couples minimally to gravity.  Its density, pressure and equation of
state are given by
\ber\label{eq:scalar}
\rho_{\phi} &=& \half \dot{\phi}^2+V(\phi), ~~
p_{\phi} = \half \dot{\phi}^2-V(\phi),\nonumber\\
w_{\phi} &=& \frac{p_{\phi}}{\rho_{\phi}} = \frac{\half \dot{\phi}^2-
V(\phi)}{\half \dot{\phi}^2+V(\phi)} \geq -1\,\,.
\eer
Scalar field evolution is governed by the equation of motion
\beq
\ddot{\phi}+3 H \dot{\phi} +\frac{dV}{d\phi} = 0~,
\eeq
where
\beq\label{eq:kinessence}
H^2 = \frac{8\pi G}{3}\left\lbrack \rho_{0m}(1+z)^3 + \half{\dot\phi}^2+
V(\phi)\right\rbrack.
\eeq
It is clear from (\ref{eq:scalar}) that $w < -1/3$ provided
$\dot{\phi}^2 < V(\phi)$. Models with this property can lead to an
accelerating universe at late times. An important subclass of
quintessence models displays the so-called `tracker' behaviour during
which the ratio of the scalar field energy density to that of the
matter/radiation background changes very slowly over a substantial
period of time.  Models belonging to this class satisfy $V^{\prime
  \prime} V/(V^{\prime})^2 \geq 1$ and approach a common evolutionary
`tracker path' from a wide range of initial conditions. As a result,
the present value of dark energy in tracker models is to a large
extent (though not entirely) independent of initial conditions and is
determined by parameters residing only in its potential -- as in the
case of the cosmological constant \citep[for a brief review of tracker
models see][]{sahni02}.  In this paper we will focus our attention on
the tracker potential $V(\phi) \propto \phi^{- \alpha}$, $\alpha \geq
1$ which was originally proposed in \citet{ratra88}. For this
potential, the region of initial conditions for $\phi$ for which the
tracker regime has been reached before the end of the matter-dominated
stage is $\phi_{in}\ll M_P\equiv 1/\sqrt G$, and the present value of
quintessence is $\phi(t_0)\sim M_P$.

For all quintessence models $w\ge -1$, and this inequality is
saturated only if $\dot \phi = dV/d\phi =0$. In order to obtain $w<-1$
matter must violate the strong energy condition $\rho + 3p \geq 0$,
for some duration of time.  It should be noted that DE with $w<-1$ is
not excluded by observations \citep[see][for a recent
investigation]{mmot02}. However in order to have $w<-1$ one must look
beyond quintessence models. Models based on scalar-tensor gravity
\citep{beps00} can have $w<-1$, so too can braneworld models (see
\citealt{ss02a} for a discussion of this issue and \citealt{as02} for a
comparison of braneworld models with observational data).

\item {\bf Chaplygin gas.}
An interesting alternate form of dark energy is provided by the
Chaplygin gas \citep[Kamenshchik \etal 2001;][]{chap2,chap3,chap4,chap5,chap6}
which obeys the equation of state
\beq
p_c = - A/\rho_c~.
\label{eq:chap_state}
\eeq 
The energy density of the Chaplygin gas evolves according to
\beq
\rho_c=\sqrt{ A+B (1+z)^6}\,\,,
\eeq
from where we see that $\rho_c \to \sqrt A$ as $z \to -1$ and $\rho_c
\to \sqrt{B}(1+z)^3$ as $z \gg 1$. Thus, the Chaplygin gas behaves
like pressureless dust at early times and like a cosmological constant
during very late times. Note, however, that Chaplygin gas at $z\gg 1$
is not simply a new kind of CDM if we examine its inhomogeneities (
\ie if we apply this hydrodynamical equation of state to the
inhomogeneous case, too)! In contrast to CDM and baryons, the sound
velocity in the Chaplygin gas $v_c= \sqrt{dp_c / d\rho_c}={\sqrt
  A / \rho_c}$ quickly grows $\propto t^2$ during the
matter-dominated stage and becomes of the order of the velocity of
light at present (it approaches light velocity asymptotically in the
distant future ).  Thus, from the point of view of inhomogeneities,
the properties of the Chaplygin gas during the matter-dominated epoch
are very unusual and resemble those of hot dark matter which has a
large Jeans length, despite the fact that the Chaplygin gas formally
carries negative pressure.
  
The Hubble parameter for a universe containing cold dark matter and
the Chaplygin gas is given by
\beq\label{eq:hub_chap} 
H(z) = H_0\left\lbrack \om(1+z)^3 + \frac{\om}{\kappa}\sqrt{\frac{A}{B} +
    (1+z)^6}\right\rbrack^{1/2}\,\,, 
\eeq 
where $\kappa = \rho_{0m}/\sqrt{B}$. 
It is easy to see from (\ref{eq:hub_chap}) that
\beq\label{eq:chap_def} 
\kappa = \frac{\rho_{0m}}{\rho_c}(z \to \infty)\,\,.  
\eeq 
Thus, $\kappa$ defines the ratio between CDM and the Chaplygin gas
energy densities at the commencement of the matter-dominated stage. It
is easy to show that
\beq 
A = B \left\lbrace \kappa^2 \left( \frac{1-\Omega_m}{\Omega_m} \right)^2 - 1
\right\rbrace \,\,.  
\eeq 
In the limiting case when $A = 0$, the Chaplygin gas becomes
indistinguishable from dust-like matter (if we examine its behaviour
in an unperturbed FRW background). This limiting case corresponds to
\beq\label{eq:chap_lim} 
\kappa = \frac{\om}{1-\om}\,\,, 
\eeq 
and is shown as the outer envelope (dashed) to the Chaplygin gas
models in Figures 1a,b. In the other limiting case $B=0$, the
Chaplygin gas reduces to the cosmological constant.

The fact that the sound velocity in the Chaplygin gas is not small
during the matter-dominated stage and becomes very large towards its
end suggests that the parameter $\kappa$ should be large in order to
avoid damping of adiabatic perturbations. This requires $A\gg B$.
Recent investigations which look at Chaplygin gas models in the light
of galaxy clustering data and CMB anisotropies show that this
observation is correct if the equation of state $p_c \propto
-1/\rho_c$ is assumed to be universally valid \citep{cf02,stzw02,bd03}. 
In our paper we consider the Chaplygin gas equation of state to be a
phenomenological description of dark energy in a FRW background and do
not assume that it remains true for perturbations. However, the fact
that $\kappa$ should be large for viable models will appear in our
results, too.  Finally let us point out that the Chaplygin gas may be
considered to be a specific case of k-essence with a constant
potential and the Born-Infeld kinetic term.  To illustrate this
consider the Born-Infeld lagrangian density
\beq
{\cal L} = -V_0\sqrt{1-\phi_{,\mu}\phi^{,\mu}}~,
\label{eq:born}
\eeq
where $\phi_{,\mu} \equiv \partial{\phi}/\partial x^\mu$.
For time-like $\phi_{,\mu}$ one can define a four velocity 
\beq
u^\mu = \frac{\phi^{,\mu}}{\sqrt{\phi_{,\alpha}\phi^{,\alpha}}}~,
\eeq
this leads to the standard form for the hydrodynamical energy-momentum tensor
\beq
T_{\mu \nu} = (\rho + p)u_\mu u_\nu - pg_{\mu \nu}~,
\eeq
where \citep{fks02}
\beq
\rho = \frac{V_0}{\sqrt{1-\phi_{,\mu}\phi^{,\mu}}}~,~~
p = - V_0\sqrt{1-\phi_{,\mu}\phi^{,\mu}}~,
\eeq
and we find that we have recovered (\ref{eq:chap_state}) with $A = V_0^2$.

\item{\bf Braneworld models.}
Braneworld models provide an interesting alternative to dark energy
model building. According to this higher dimensional world view, we
live on a 3+1 dimensional brane (`brane' being a multidimensional
generalization of `membrane') which is either embedded in or bounds a
higher dimensional space-time.  The simplest example of a braneworld
which can lead to late-time acceleration is the model suggested by
Deffayet \etal (2002) (we shall henceforth refer to this model as the
DDG model).
\beq\label{eq:ddg}
H = \sqrt{\frac{8\pi G \rho_{\rm m}}{3} + \frac{1}{l_c^2}} + 
\frac{1}{l_c}~,
\eeq
where $l_c = m^2/M^3$ is a new length scale and $m$ and $M$ refer
respectively to the four and five dimensional Planck mass ($l_c =
2r_c$ in the terminology of Deffayet \etal 2002).  The acceleration of
the universe in this model is not caused by the presence of `dark
energy' but due to the fact that general relativity is formulated in 5
dimensions instead of the usual 4.  One consequence of this is that
gravity becomes five dimensional on length scales $R > l_c =
2H_0^{-1}(1-\Omega_{\rm m})^{-1}$.  A more general class of braneworld
models is described by \citep{ss02a}
\ber \label{eq:hubble_brane}
{H^2(z) \over H_0^2} &=& \Omega_{\rm m} (1\!+\!z)^3 + \Omega_\sigma +\underline{2 \Omega_l \mp} \nonumber\\
&&\underline{2 \sqrt{\Omega_l}\,
\sqrt{\Omega_{\rm m} (1\!+\!z )^3 + \Omega_\sigma + \Omega_l +
\Omega_{\l_{\rm b}}}} \, ,
\eer
where $\l_{\rm b}$ is the bulk cosmological constant, $\sigma$ is the
brane tension and
\beq \label{eq:omegas}
\Omega_{\rm m} =  {\rho_{0m} \over 3 m^2 H_0^2} , \Omega_\sigma 
= {\sigma \over 3 m^2H_0^2} ,
\Omega_l = {1 \over l_c^2 H_0^2} , \Omega_{\l_{\rm b}} = - {\l_{\rm b} \over 6 H_0^2} 
\eeq
It is easy to see that $l_c$ can be of the same order as the Hubble
radius $l_c \sim H_0^{-1}$ if $M \sim 100$ MeV.  On short length
scales $r \ll l_c$ and at early times, one recovers general
relativity, whereas on large length scales $r \gg l_c$ and at late
times brane-related effects begin to play an important role.  It is
interesting that brane-inspired effects can lead to the late time
acceleration of the universe even in the complete absence of a matter
source which violates the strong energy condition $\rho+3p \geq 0$
\citep[Deffayet \etal 2002;][]{ss02a}.

The dimensionless value of the brane tension $\Omega_\sigma$ is determined 
by the constraint relation
\beq \label{eq:omega1}
\Omega_{\rm m} + \Omega_\sigma \mp \underline{2
\sqrt{\Omega_l}\, \sqrt{1 - \Omega_\kappa + \Omega_{\l_{\rm b}}}} = 1.
\eeq

The underlined terms in (\ref{eq:hubble_brane}) \& (\ref{eq:omega1})
make braneworld models
different from standard FRW cosmology. Indeed by setting $\Omega_l = 0$
(\ref{eq:hubble_brane}) reduces to the LCDM model
\beq
{H^2(z) \over H_0^2} = \Omega_{\rm m} (1\!+\!z)^3 + \Omega_\sigma
\eeq
which describes a universe containing matter and a cosmological
constant (\ref{eq:lambda}).  The two signs in (\ref{eq:hubble_brane})
correspond to the two separate ways in which the brane can be embedded
in the higher dimensional bulk.  As shown in \citet{ss02a}, taking the
upper sign in (\ref{eq:hubble_brane}) and (\ref{eq:omega1}) leads to
the model called BRANE1, while the lower sign in
(\ref{eq:hubble_brane}) and (\ref{eq:omega1}) results in BRANE2.

Three important classes of braneworld models deserve special mention:
\begin{enumerate}

\smallskip
\item BRANE1 models have an effective equation of state which is {\em
  more negative} than that of the cosmological constant $w \leq -1$.

\smallskip
\item BRANE2 models have $w \geq -1$.  For parameter values
$\Omega_\sigma = \Omega_{\l_{\rm b}}=0$, BRANE2 coincides with the
dark energy model discussed in Eq. (\ref{eq:ddg}).

\smallskip
\item A class of braneworld models, called `disappearing dark energy' 
(DDE) \citep{ss02a,as02}, have the important property that the 
current acceleration of the universe is a {\em transient phase} 
which is sandwiched between two matter dominated epochs. These 
models do not have horizons and therefore help to reconcile an 
accelerating universe with the demands of the string/M-theory
\citep{sahni02} (as well as any theory which requires dark energy to
decay in the future and transform into matter with $w\ge -1/3$).
\end{enumerate}

Finally we note that, for a spatially flat universe, the luminosity
distance for all models discussed above is given by the simple
expression
\beq\label{eq:lumdis}
\frac{D_L(z)}{1+z} =  \int_0^{z} \frac{dz'}{H(z')}
\eeq
where $H(z)$ is given by (\ref{eq:lambda}) for LCDM, by
(\ref{eq:quiessence}) for quiessencence, by (\ref{eq:kinessence}) for
quintessence, by (\ref{eq:hub_chap}) for the Chaplygin gas and by
(\ref{eq:hubble_brane}) for the braneworld models.
\end{itemize}

\begin{figure*}
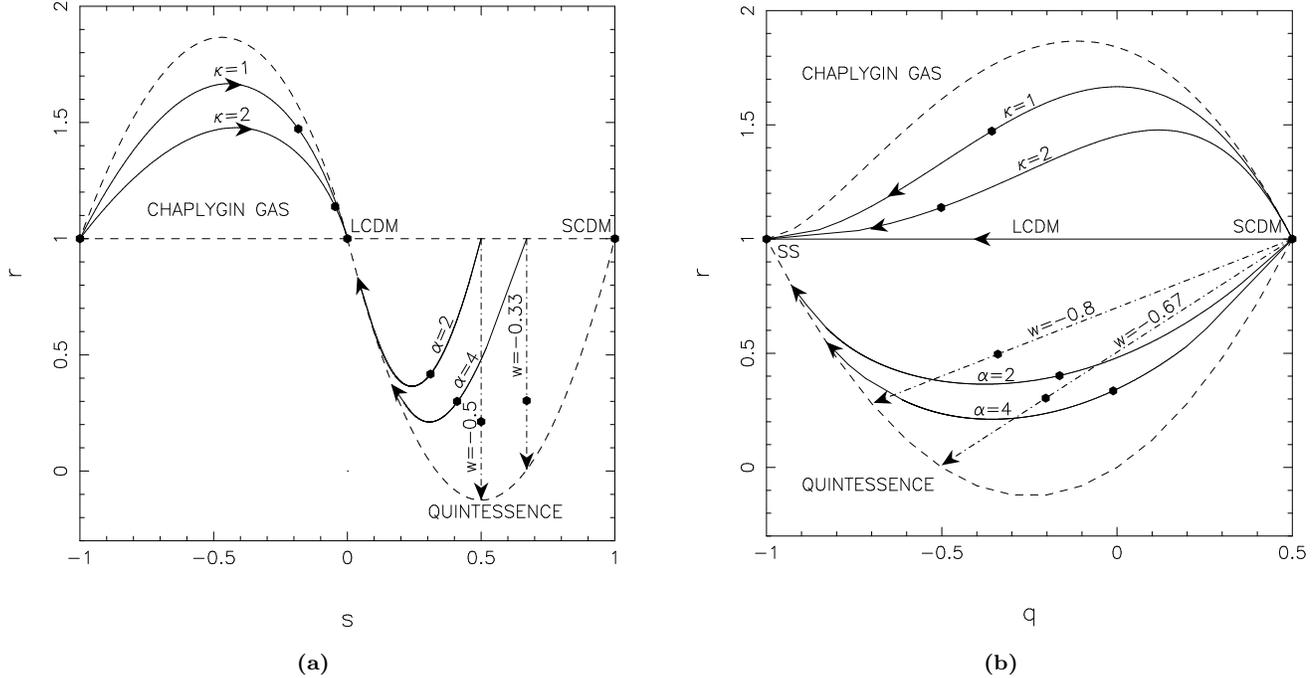
 \centering
\begin{center}
$\begin{array}{c@{\hspace{0.4in}}c}
\multicolumn{1}{l}{\mbox{}} &
        \multicolumn{1}{l}{\mbox{}} \\ [0.0cm]
\epsfxsize=3.2in
\epsffile{evolve_rs.epsi} &
        \epsfxsize=3.2in
        \epsffile{evolve_qr.epsi} \\ [0.20cm]
\mbox{\bf (a)} & \mbox{\bf (b)}
\end{array}$
\end{center}
\caption{\small 
The left panel (a) shows the time evolution of the statefinder pair
$\lbrace r,s \rbrace$ for quintessence models and the Chaplygin gas.
Quintessence models lie to the right of the LCDM fixed point
($r=1,s=0$) (solid lines represent tracker potentials
$V=V_0/\phi^{\alpha}$, dot-dashed lines representing quiessence with
constant equation of state $w$).  For quiessence models, $s$ remains
constant at $1+w$ while $r$ declines asymptotically to
$1+\frac{9}{2}w(1+w)$. For tracker models, $s$ monotonically decreases
to zero whereas $r$ first decreases from unity to a minimum value,
then rises to unity. These models tend to approach the LCDM fixed
point ($r=1, s=0$) from the right at $t \to \infty$.  Chaplygin gas
models (solid lines) lie to the left of the LCDM fixed point. For
these models, $\kappa$ is the ratio between matter density and the
density of the Chaplygin gas at early times.  For all Chaplygin gas
models, $s$ monotonically increases to zero from -1, whereas $r$ first
increases from unity to a maximum value, then decreases (to unity).
The dashed curve in the lower right is the envelope of all
quintessence models, while the dashed curve in the upper left is the
envelope of Chaplygin gas models (the latter is described by $\kappa =
\om/1-\om$).  The region outside the dashed curves is forbidden for
both classes of dark energy models.  The right panel (b) shows the
time evolution of the pair $\lbrace r,q\rbrace$, where $q$ is the
deceleration parameter. It is important to note that the the solid
line, which corresponds to the time evolution of the LCDM model,
divides the $r-q$ plane into two halves. The upper half is occupied by
Chaplygin gas models, while the lower half contains quintessence
models. All models diverge at the same point in the past ($r=1,q=0.5$)
which corresponds to a matter dominated universe (SCDM), and converge
to the same point in the future ($r=1,q=-1$) which corresponds to the
steady state model (SS) -- the de Sitter expansion (LCDM $\to$ SS as
$t \to \infty$ and $\om \to 0$).  The dark dots on the curves show
current values $\lbrace r_0, s_0\rbrace$ (left) and $\lbrace r_0,
q_0\rbrace$ (right) for different dark energy models. In all models,
$\om = 0.3$ at the current epoch.  In both panels quiessence is shown
as dot-dashed while dashed lines mark envelopes for Chaplygin gas
(upper) and quintessence (lower).  }
\label{fig:evolve}
\end{figure*}

\section{The statefinder diagnostic}
\label{sec:statefinder}

As we have seen above, dark energy has properties which can be {\em
  very} model dependent.  In order to be able to differentiate between
the very distinct and competing cosmological scenarios involving dark
energy, a sensitive and robust diagnostic (of dark energy) is a must.
Although the rate of acceleration/deceleration of the universe can be
described by the single parameter $q = -\ddot{a}/a H^2$, a more
sensitive discriminator of the expansion rate and hence dark energy
can be constructed by considering the general form for the expansion
factor of the Universe
\beq\label{eq:taylor0}
a(t) = a(t_0) + {\dot a}\big\vert_0 (t-t_0) + 
\frac{\ddot{a}\big\vert_0}{2} (t-t_0)^2 + 
\frac{\atridot\big\vert_0}{6} (t-t_0)^3 + ...~.
\eeq

In general, dark energy models such as quiessence, quintessence,
k-essence, braneworld models, Chaplygin gas etc. give rise to families
of curves $a(t)$ having vastly different properties.  Since we know
that the acceleration of the universe is a fairly recent phenomenon
\citep{benitez,riess01,ss00} we can, in principle, confine our
attention to small values of $\vert t-t_0\vert$ in (\ref{eq:taylor0}).
We have shown in \citet{sssa02} that a new diagnostic of dark energy
called statefinder can be constructed using both the second and third
derivatives of the expansion factor.  The second derivative is encoded
in the deceleration parameter which has the following form in a
spatially flat universe:
\beq\label{eq:decel1} 
q = - \frac{\ddot a}{aH^2} \equiv \frac{1}{2} \left (1 + 3w\omx\right )~, ~~~\Omega_X=1-\Omega_m~.  
\eeq
The statefinder pair $\statei$, defines two new cosmological
parameters (in addition to $H$ and $q$):
\ber\label{eq:statefinder1}
r &\equiv& \frac{\atridot}{a H^3} = 1 + \frac{9w}{2} \omx(1+w) -
\frac{3}{2} \omx \frac{\dot{w}}{H} \,\,,\\ 
s &\equiv& \frac{r-1}{3(q-1/2)} = 1+w-\frac{1}{3} \frac{\dot{w}}{wH}\,\,.
\eer

Clearly an important requirement of any diagnostic is that it permits
us to differentiate between a given dark energy model and the simplest
of all models -- the cosmological constant $\l$. The statefinder does
exactly this. For the LCDM model, the value of the first statefinder
stays pegged at $r=1$ even as the matter density evolves from a large
initial value ($\om \simeq 1, ~t \ll t_0$) to a small late-time value
($\om \to 0, ~t \gg t_0$). It is easy to show that $\statei = \lbrace
1,0\rbrace$ is a fixed point for LCDM.

The second statefinder $s$ has properties which complement those of
the first. Since $s$ does not explicitly depend upon either $\omx$ or
$\om$, many of the degeneracies which are present in $r$ are broken in
the combined statefinder pair $\statei$.  For models with a constant
equation of state (quiessence) $s = 1+w = $ constant, while the
statefinder $r$ is time-varying. For models with time-dependent
equation of state (kinessence), both $r$ and $s$ vary with time.  As
we will show in this paper, the statefinder pair $\statei$ can easily
distinguish between LCDM, quiessence and kinessence models. It can
also distinguish between more elaborate models of dark energy such as
braneworld models and the Chaplygin gas \citep[see also][Gorini \etal
2002]{sssa02}. Interestingly, as demonstrated in section
\ref{sec:results}, the statefinder pair $\statek$ proves to be an even
better diagnostic of dark energy than $\statei$.

\begin{figure} 
\centering
\centerline{ \psfig{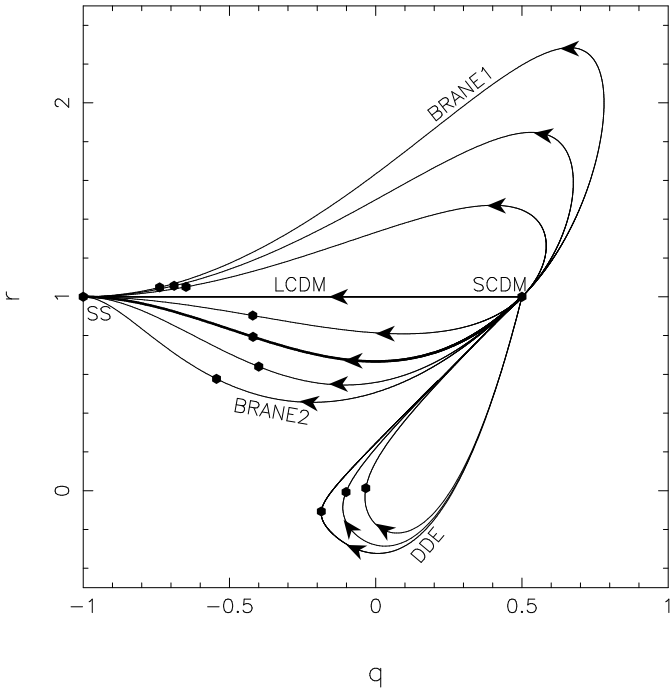} }
\caption{\small
Trajectories in the statefinder plane $\statej$ for the braneworld
models discussed in (\ref{eq:hubble_brane}).  BRANE1 models have $w
\leq -1$ generically, whereas BRANE2 models have $w \geq -1$.  The
closed loop represents DDE in which the acceleration of the universe
is a transient phenomenon.  For braneworld models, parameter values
are as follows.  BRANE1: solid curves above LCDM; top to bottom:
$\om=0.6, \Omega_l=6.0$, $\om =0.5, \Omega_l=2.0$, $\om=0.4,
\Omega_l=0.5$.  BRANE2: solid curves below LCDM; top to bottom:
$\om=0.3, \Omega_l=0.05$, $\om=0.2, \Omega_l=0.25$, $\om=0.1,
\Omega_l=0.45$.  The thick solid curve in BRANE2 corresponds to the
DDG model discussed in (\ref{eq:ddg}) with $\om = 0.24$.  For BRANE1
and BRANE2 models $\Omega_{\Lambda_b} = 0$ (\ie there is no
cosmological constant in the bulk.)  Dark dots indicate the current
value of $\statej$ for the models.  All models are in reasonable
agreement with current supernova data.  For DDE, from outer to inner
loops, $\om=0.05, \Omega_{\Lambda_b}=4.9$, $\om=0.15,
\Omega_{\Lambda_b}=1.4$, $\om=0.20, \Omega_{\Lambda_b}=1.1$.  }
\label{fig:brane}
\end{figure}

The  statefinders $r$ and $s$ can be easily expressed in terms of the 
Hubble parameter $H(z)$ and its derivatives as follows:
\ber\label{eq:statefinder}
r(x) &=& 1-2 \frac{H^{\prime}}{H} x+\left\lbrace\frac{H^{\prime\prime}}{H}
+\left(\frac{H^{\prime}}{H}\right)^2\right\rbrace x^2,\nonumber\\
s(x) &=& \frac{r(x)-1}{3 ( q(x)-1/2)}\,\,,
\eer
where $x=1+z$ and $H$ is given by (\ref{eq:lambda}), (\ref{eq:quiessence}),
(\ref{eq:kinessence}), (\ref{eq:hub_chap}), (\ref{eq:hubble_brane})
for the different dark energy models discussed in the previous section.

In figure~\ref{fig:evolve}(a) we show the time evolution of the
statefinder pair $\statei$. We find that the vertical line at $s=0$
effectively divides the $r-s$ plane into two halves. The left half
contains Chaplygin gas (CG) models which commence their evolution from
$r=1, s=-1$ and end it at the LCDM fixed point ($r=1, s=0$) in the
future.  The quintessence models occupy the right half of the $r-s$
plane. These models commence their evolution from the right of the
LCDM fixed point and, like CG, are also attracted towards the LCDM
fixed point in the future.  For quiessence models, $r$ decreases
monotonically to $1+\frac{9}{2}w(1+w)$ while $s$ remains constant at
$s=1+w$. For kinessence models, on the other hand, $s$ decreases
monotonically to zero, while $r$ first decreases to a minimum value
then increases to unity. The region below the curve
$r=1+\frac{9}{2}s(s-1)$ is disallowed for quintessence models whereas
CG models with $\kappa < \om/(1-\om)$ are excluded.  It is interesting
that the second statefinder, $s$, is positive for quintessence models,
but negative for the CG.  Similarly the first statefinder, $r$, is $<
1$ ($> 1$) for quintessence (CG). The distinctive trajectories which
quiessence, quintessence and CG follow in the $r-s$ plane demonstrates
quite strikingly the contrasting behaviour of dark energy models.

The separation between distinct families of dark energy models is also
very pronounced when we analyze evolutionary trajectories using the
statefinder pair $\statej$ shown in Fig.~\ref{fig:evolve}(b) and
Fig.~\ref{fig:brane}. Fig.~\ref{fig:evolve}(b) shows the evolution of
quintessence and CG models in $r-q$ space, while Fig.~\ref{fig:brane}
shows the evolution of the braneworld models discussed in
(\ref{eq:hubble_brane}). In Fig.~\ref{fig:evolve}(b) the LCDM model
effectively divides the $r-q$ space in half, separating quintessence
models (bottom-half) from the Chaplygin gas (top-half).  From this
figure we clearly see that all dark energy models commence evolving
from the same point in the past ($q=0.5, r=1$), which corresponds to a
matter dominated SCDM universe. Quintessence, LCDM and the Chaplygin
gas all end their evolution at the same common point in the future
($q=-1, r=1$), which corresponds to steady state cosmology (SS) -- the
de Sitter expansion.  In Fig.~\ref{fig:brane} the LCDM model separates
BRANE1 models (which have $w_{\rm eff} \leq -1$) from BRANE2 models as
well as DDE models.  BRANE2 models have $w_{\rm eff} \geq -1$
generically, whereas DDE models consist of a transient accelerating
regime which is sandwiched between two matter dominated epochs. Thus
DDE both begins and ends its evolution at the SCDM point $\statej =
\lbrace 1, 0.5 \rbrace$ and its $r-q$ space trajectory is a loop !
BRANE1 and BRANE2 models on the other hand, commence evolving at the
SCDM point and tend to SS in the future.  Fig. \ref{fig:evolve}(b) and
Fig. \ref{fig:brane} clearly demonstrate that the deceleration
parameter cannot on its own differentiate between rival models of dark
energy. The degeneracy which afflicts $q(z)$ clearly also afflicts the
equation of state $w(z)$, since both $q$ and $w$ are related through
(\ref{eq:state}).  We therefore feel we have convincingly demonstrated
that the statefinders can successfully differentiate between competing
dark energy models as diverse as LCDM, quintessence, braneworld models
and the Chaplygin gas. Statefinders can also be applied to other
interesting candidates for dark energy including bigravity models
\citep{damour02}, generalized Chaplygin gas \citep{chap1,bento02},
k-essence \citep{mukhanov} scalar-tensor theories etc.

Finally we draw the readers attention to the following elegant
relationship which exists between the statefinders on the one hand,
and the {\rm total density} $\rho = \sum_i \rho_i$ and {\rm total
  pressure} $p = \sum_i p_i$ in the universe:
\beq\label{eq:statefinder_new}
q-{1\over 2} = {3p\over 2\rho}~,~~
r-1 ={9(\rho + p)\over 2\rho}{\dot{p}
\over
\dot{\rho}}~,~~s={(\rho + p)\over p}
{\dot {p} \over \dot{\rho}}~.
\eeq
From Eq. (\ref{eq:statefinder_new}) we see that the statefinder $s$ is
exceedingly sensitive to the total pressure $p$.  This has some
interesting consequences.  At early times the presence of radiation
ensures that the total pressure in the universe is positive. Much
later, the universe begins to accelerate driven by the negative
pressure of dark energy. In between these two asymptotic regimes, deep
in the matter dominated epoch, a stage is reached when the (negative)
pressure of dark energy is exactly balanced by the positive pressure
of radiation. At this precise moment of time $p \simeq 0$ and $s \to
\infty$ !  For LCDM this pressure balance is achieved at $z_* \sim
10$, consequently $\vert s \vert \gg 1$ when $z \sim z_*$. It can be
shown that the redshift $z_*$ (at which $p = 0$) is quite sensitive to
the form of dark energy. We therefore find that the statefinder $s$
diagnoses the presence of dark energy even at high redshifts when the
contribution of DE to the total energy budget of the universe is
insignificant !

\section{Model independent reconstruction of cosmological parameters from 
SNAP data}

\subsection{The cosmological reconstruction of dark energy properties}
\label{sec:cosmo_fits}
Cosmological reconstruction is an effective statistical technique
which can be used in situations where a large number of theoretical
models are to be compared with observations. Instead of estimating
relevant parameters for each model separately, we can choose a
model-independent fitting function and perform a maximum likelihood
parameter estimation for it. The resultant confidence levels can be
used to rule out or accept the different models available. This
technique is effective here because, as discussed in
Section~\ref{sec:darkenergy}, a wide range of theoretical models have
been suggested to explain dark energy.

The basis of cosmological reconstruction rests in the observation that
the expression for the luminosity distance (\ref{eq:lumdis}) can be
easily inverted \citep{st00,turner,nak99}: \beq\label{eq:H}
H(z)=\left[{d\over dz}\left({D_L(z)\over 1+z}\right)\right]^{-1}~.
\eeq Thus, from mathematical point of view, any given $D_L(z)$ defines
$H(z)$. Eqs. (\ref{eq:energy}) and (\ref{eq:state}) can then be used
to obtain the dark energy density and the associated equation of
state.  Similarly the statefinder pair $\statei$ can be determined by
employing Eq (\ref{eq:statefinder}) together with Eq
(\ref{eq:decel1}).  However, in practice the derivative with respect
to $z$ may not be simply performed since $D_L(z)$ is noisy due to
observational errors (mainly, due to variance in supernovae
luminosity). Therefore, the {\em smoothing} of data over some interval
$\Delta z$ is required ($\Delta z$ may depend on $z$). The value of
$\Delta z$ is determined by estimated errors and by the required
accuracy with which we want to determine $H(z)$. Of course, the
resulting $H(z)$ will be smoothed, too, as compared to the genuine
one. Note that our presentation here is very similar to that in
\citet{teg02}.

Instead of actually dividing a measured range of $z$ into intervals,
one may {\em parametrize} $H(z)$ by some fitting curve which depends
on a number of free parameters. This leads to model-independent
parametric reconstruction of $H(z)$, $\rho_{DE}(z)$, $w_{\rm eff}(z)$
and other quantities. It is clear that the number of free parameters
$N$ in such a fit just defines the equivalent smoothing interval
$\Delta z$ (in particular, $\Delta z = z_{max}/N$ if $\Delta z$ is
chosen to be independent of $z$ and we are considering the function
$H(z)/H_0$, so that its value at $z=0$ is known exactly). Thus, the
parametrization is equivalent to some kind of smoothing, with the
actual way of smoothing (weighting) depending on the functional form
of the parametric fit used. This refers even to such sophisticated
methods as the `principal-component' approach used in \citet{hut02}.
Since decreasing $\Delta z$ (increasing $N$) results in a rapid growth
of errors ($\Delta H(z) \propto (\delta z)^{-3/2}$ directly follows
from Eq. (\ref{eq:H}), c.f. \citet{teg02}), for a given $z_{max}$
there is no sense in taking $N$ to be large -- this will merely result
in the loss of accuracy of our reconstruction. Thus, we will consider
only 3-parametric fits for $H(z)$ (these will correspond to
2-parametric fits for $w(z)$).

After the discovery that the universe is accelerating, many different
fitting function approaches were suggested and some are summarized
below.

\begin{itemize}

\item{\bf Polynomial Fit to Dark Energy : }

In this paper, we reconstruct dark energy using a very effective ansatz 
introduced in \citet{sssa02} in which the
dark energy density
is expressed as a truncated Taylor series polynomial in $x=1+z$, 
$\rho_{\rm DE} = A_1 + A_2x + A_3 x^2$. 
This leads to the following ansatz for the Hubble parameter 
\beq\label{eq:poly_fit}
H(x) =  H_0\left\lbrack \om x^3 + A_1 + A_2x + A_3 x^2\right\rbrack^\half\,\, ,
\label{eq:taylor}
\eeq
which, when substituted in the expression for the luminosity distance 
(\ref{eq:lumdis}), yields
\beq
\frac{D_L}{1+z} =  \frac{c}{H_0}\int_1^{1+z} \frac{dx}{\sqrt{\om x^3 + 
A_1 + A_2 x + A_3 x^2}}\,\,.
\label{eq:taylor1}
 \eeq
 
The values of the parameters $A_1, A_2, A_3$ are obtained by fitting
(\ref{eq:taylor1}) to supernova observations by means of a maximum
likelihood analysis discussed in the next section.  There are obvious
advantages in choosing the ansatz (\ref{eq:poly_fit}) namely, it is
exact for the cosmological constant $w = -1$ ($A_2 = A_3 = 0$) as
well as for quiessence with $w = -2/3$ ($A_1 = A_3 = 0$) and $w =
-1/3$ ($A_1 = A_2 = 0$).  Furthermore, the presence of the term $\om
x^3$ in (\ref{eq:poly_fit}) ensures that the ansatz correctly
reproduces the matter dominated epoch at early times ($z \gg 1$).
The presence of this term also allows us to incorporate information
pertaining to the value of the matter density and, as we shall soon
demonstrate, permits elaborate statistical analysis with the
introduction of priors on $\om$.

The statefinder pair for the polynomial fit (\ref{eq:poly_fit}) can be
written in terms of $x=1+z$ as follows 
\ber\label{eq:statefinder3}
r(x) &=& \frac{\om x^3+A_1}{\om x^3+A_1+A_2 x+A_3 x^2} ,\\
s(x) &=& \frac{2 (A_2 x+A_3 x^2)}{3 (3 A_1+2 A_2 x+A_3 x^2)} \,\,.
\eer

It is also straightforward to obtain expressions for the
cosmological parameters $q$ and $w$ by substituting (\ref{eq:taylor})
in (\ref{eq:decel}) and (\ref{eq:state}) respectively.

\begin{figure} 
\centering 
\centerline{ \psfig{figure=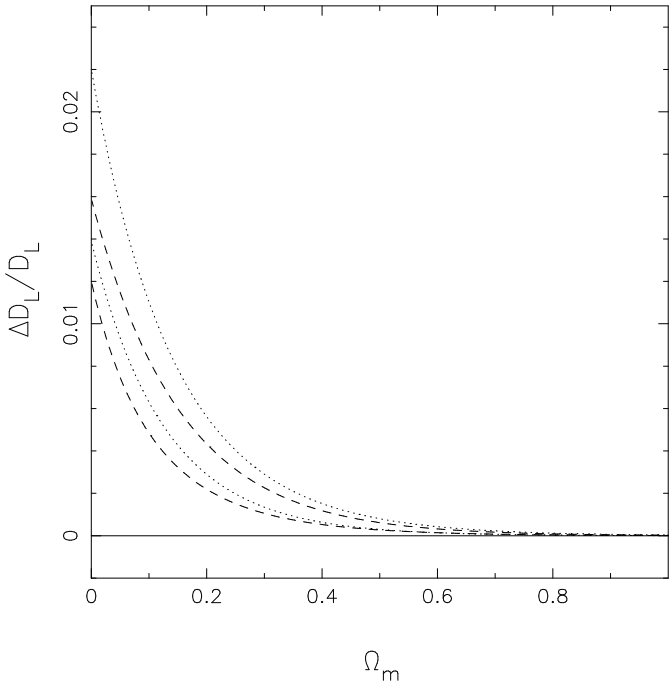,width=0.5\textwidth,angle=0} }
\caption{\small
The maximum deviation $\vert\Delta D_L/D_L\vert$ between the actual
value of the luminosity distance in the redshift range $z=0-10$ in a
DE model and that calculated using the {\em polynomial fit}
Eq~(\ref{eq:taylor1}).  The solid line at $\Delta D_L/D_L=0$
represents models with $w=-1, w=-2/3, w=-1/3$, for which the
polynomial fit returns exact values . The dashed lines from top to
bottom represent the tracker potential $V(\phi) = V_0/ \phi^{\alpha}$
for $\alpha=1 \ {\rm and} \ 2$ respectively. The dotted lines
represent Chaplygin gas models with $\kappa=0.5 \ {\rm and} \ 2$ (top
to bottom).}
\label{fig:h_exp_err.ps}
\end{figure}

In figure~\ref{fig:h_exp_err.ps} we show the {\em maximum deviation}
between the exact value of the luminosity distance and the
fit-estimated approximate value for a class of dark energy models.
For LCDM ($w = -1$) and two quiessence models ($w = -2/3$, $w =
-1/3$), the ansatz (\ref{eq:taylor1}) returns {\em exact} values. (The
ansatz is also exact for SCDM.)  For the two tracker and Chaplygin gas
models which we consider, the luminosity distance is determined to
better than 1\% accuracy for a conservative range in $\om$ ($0.2 \leq
\om \leq 0.5$).  We therefore conclude that the polynomial fit
(\ref{eq:taylor1}) is very accurate and can safely be applied to
reconstruct the properties of dark energy models.

In this paper we will use the polynomial fit (\ref{eq:taylor1}) to
perform a model independent reconstruction of dark energy using the
synthetic SNAP supernova data discussed earlier.  Some details of our
approach which involves the maximum likelihood method will be
discussed in sections \ref{sec:ml}.  Our results for the cosmological
reconstruction of dark energy using the statefinder will be presented
in section \ref{sec:results}.

Although we will mainly work with the polynomial ansatz (\ref{eq:taylor}) 
to reconstruct the properties of the statefinders,
it is worthwhile to summarize some of the alternate approaches to the
cosmological reconstruction problem.

\item{\bf Fitting functions to the luminosity distance $D_L$ : } An
interesting complementary approach to the reconstruction exercise is
to find a suitable fitting function for the luminosity distance.
Such an approach was advocated in \citet{turner} and \citet{srss00}.
In \citet{turner} a polynomial fit for the luminosity distance was
suggested which had the form
\beq
\frac{D_L(z)}{1+z} = \sum_{i=1}^N a_iz^i\,\,.
\label{eq:poly}
\eeq
The ansatz (\ref{eq:poly}) was examined in \citet{albrecht} who
demonstrated that this approximation does not give an accurate
reconstruction of the equation of state of dark energy. Similar
conclusions will also be reached by us later in this paper in
connection with the reconstruction of the statefinder pair using
(\ref{eq:poly}).

A considerably more versatile and accurate fitting function to the 
luminosity distance is 
\citep{srss00}
\beq               
{D_L\over x} 
\equiv \frac{2}{H_0}\left[ \frac{x - A_1\sqrt{x} -1 + A_1}{A_2 x+ 
A_3\sqrt{x} + 2 - A_1 -A_2 -A_3}\right] ~, ~~ x = 1+z~,      
\label{eqn:star}  
\eeq
where $A_1$, $A_2$ and $A_3$ are parameters whose values must be
determined by fitting (\ref{eqn:star}) to observations. Important
properties of this function are that it is valid for a wide range of
models and that it {\it exactly reproduces} the results both for SCDM
($\Omega_m = 1$) and the steady state model ($\Omega_\l=1$).  As
demonstrated in \citet{srss00}, an accurate analytical form for $D_L$
allows us to reconstruct the Hubble parameter by means of the relation
(\ref{eq:H}).  Cosmological parameters including $q(z)$, $w(z)$,
$r(z)$, $s(z)$ can now be easily reconstructed using (\ref{eq:decel}),
(\ref{eq:state}) and (\ref{eq:statefinder}).

\item{\bf Fitting functions to the Equation of State : } 
  
A somewhat different approach fits the equation of state of dark
energy by the first few terms of a Taylor series expansion
\citep{albrecht}:
\beq
w_{\rm DE}(z) = \sum_{i=0}^N w_iz^i\,\,.
\label{eq:wpol}
\eeq

For $N=1$ the luminosity distance can be expressed as 
\ber
\frac{D_L}{1+z} &=& \frac{c}{H_0} \int_1^{1+z} \frac{dx}{\sqrt{\om x^3 + \omx}}\,\,,\nonumber\\
\omx &=& (1-\om) x^{3 (1+w_0)} \ {\rm exp} \lbrace 3 w_1(x-1) \rbrace \,\,.
\eer

A modification of the above prescription was suggested in
\citet{efstathiou} which used a logarithmic expansion of the equation
of state of dark energy:

\beq
w(z) = w_0-\alpha \ {\rm ln} (1+z)\,\,,
\label{eq:wlog}
\eeq
where $\alpha=dw/d({\rm ln} a)$.  Yet another approach \citep{maor}
advocated a quadratic fit to the {\em total} equation of state:
\beq
w_T(z) = w_0+w_1 z+w_2 z^2\,\,, 
\label{eq:wtot}
\eeq
where the total equation of state, $w_T(z)$, is defined in terms of
the equation of state of dark energy, $w(z)$, as
\beq
w_T(z) = \frac{w(z)}{1+\frac{\om}{1-\om}{\rm exp}\left[-3 \int_1^{1+z} w(x)\frac{dx}{x}\right]}\,\,.
\eeq

\end{itemize}

Other approaches to the reconstruction problem can be found in
\citet{chiba,copeland,linder}.

\subsection{Maximum Likelihood Estimation of cosmological parameters \label{sec:ml}} 
In order to determine how effective the statefinders are in
discriminating between dark energy models, we adopt the method of
maximum likelihood estimation to our reconstruction exercise.
Supernova data is expected to improve greatly over the next few years.
This improvement will be spurred by ongoing efforts by the Supernova
Cosmology Project \footnote{http://www-supernova.lbl.gov} and the
High-z supernova search team
\footnote{http://cfa-www.harvard.edu/cfa/oir/Research/supernova/HighZ.html},
as well by planned surveys such as the Nearby SN Factory
\footnote{http://snfactory.lbl.gov} (300 SNe at $z \lleq 0.1$) and the
SuperNova Acceleration Probe -- SNAP
\footnote{http://snfactory.lbl.gov} ($\sim 2000$ SNe at $z \lleq
1.7$).  We shall use data simulated according to the specifications of
SNAP -- a space based mission which is expected to greatly increase
both the number of Type Ia SNe observed and the accuracy of SNe
observations.

{\bf SuperNova Acceleration Probe (SNAP)}

The SNAP mission is expected to observe about 2000 Type Ia SNe each
year, over a period of three years, according to the specifications
given in Table~\ref{tab:SNAP}.  We assume a Gaussian distribution of
uncertainties and an equidistant sampling of redshift in four redshift
ranges. The errors in the redshift are of the order of $\delta
z=0.002$. The statistical uncertainty in the magnitude of SNe is
assumed to be constant over the redshift range $0 \leq z \leq 1.7$ and
is given by $\sigma_{\rm mag} = 0.15$. The systematic uncertainty
limit is $\sigma_{sys}=0.02$ mag at redshift $z=1.5$. For simplicity
we assume a linear drift from $\sigma_{sys}=0$ at $z=0$ to
$\sigma_{sys}=0.02$ at $z=1.5$, so that the systematic uncertainty on
the model data is given by $\sigma_{sys}(z)=(0.02/1.5) z$.

Optimizing the model with $2000$ data points is somewhat time
consuming therefore we produced a smaller number of binned SNe
luminosity distances by binning the data in a redshift interval
$\Delta z = 0.02$. This interval is comparable to the statistical
uncertainty in the redshift measurement of high-$z$ SNe due to the
peculiar velocities of the galaxies in which they reside, which is
typically of the order of $v_{\rm peculiar} \approx 1000\,\,{\rm
  Km}\,{\rm s}^{-1} $. In our experiment we smoothed the data in the
first three redshift intervals in Table~\ref{tab:SNAP} by binning, the
last interval had relatively fewer Sne and was left unbinned.  The
statistical error in magnitude, and hence in the luminosity distance
is weighed down by the factor $1/\sqrt{N_{\rm bin}}$, where $N_{\rm
  bin}$ is the number of SNe in each bin.

\begin{table*}
\centering
\caption{Expectations from SNAP for a single year period of observation}
\label{tab:SNAP}
\begin{center}
\begin{tabular}{ccccc} \hline
{Redshift Interval}  & $z=0$--$0.2$ &  $z=0.2$--$1.2$ & $z=1.2$--$1.4$
& $z=1.4$--$1.7$   \\\hline
{Number of SNe}  & $50$ &  $1800$ & $50$
& $15$ \\\hline
\end{tabular}
\end{center}
\end{table*}

We use SNAP specifications to construct mock SNe catalogues. We may
then use the method of maximum likelihood parameter estimation on this
mock data to estimate the different cosmological parameters of
interest.

{\bf Maximum Likelihood Estimation: } 

The observable quantity for a given supernova is its bolometric or
`apparent' magnitude $m$ which is a measure of the light flux received
by us from the supernova.  To convert from $m$ to cosmological
distance, we use the well known relationship between the luminosity
distance $D_L$ and the bolometric magnitude
\beq
 m = M_0 + 25 + 5\,\log  D_L,
\label{eq:mz}
\eeq
where $M_0$ is the absolute magnitude of the SNe and the luminosity
distance $D_L$ is measured in the units of Mpc.  (For Type Ia SNe, the
typical apparent magnitude at $z=1$ is about $25$, which shows that we
are dealing with very faint objects at that redshift.)  Type Ia
supernovae are excellent standard candles, and the dispersion in their
apparent magnitude is $\sigma_{\rm mag} = 0.15$, which is nearly
independent of the SN redshift.  To relate this to the dispersion in
the measured luminosity distance, we use Eq.~(\ref{eq:mz}) to obtain
\beq
\frac{\sigma_{\rm dist}}{D_L} = \frac{\ln 10}{5} \sigma_{\rm mag} = 0.069\,\,.
\eeq 
While constructing mock SNe catalogues we shall assume that the errors
in the luminosity distance are Gaussian with zero mean and dispersion
given by the above expression ($\sim 7\%$), the normalized likelihood
function is therefore given by
\ber
L(y_i,p_k) &=&  \prod_{i=1}^{N_{\rm dat}}  \left (\frac{1}{\sqrt{2 \pi} \sigma_{\rm dist}(z_i)} \right)  \nonumber\\ 
&& \times \exp \left [ -\frac{1}{2} \left ( \frac{ y_i - D_L^{\rm fit}
 (z_i; p_i)}{\sigma_{\rm dist}(z_i)} \right )^2 \right ]\,\,,
\label{eq:lhood}
\eer
where the index $i$ ranges from $1$ to $N_{\rm dat}$, which is the
number of supernovae in our sample, and we have denoted the fiducial
supernova luminosity distance at a redshift $z=z_i$ as $y_i \equiv
D_L(z_i)$, where $D_L(z)$ is the luminosity distance simulated with
SNAP specifications for a chosen background model using the
(\ref{eq:lumdis}). The $p_i$'s are the parameters of the fitting
function.  (We shall mostly exploit the fitting function
(\ref{eq:taylor1}) for which $p_i \equiv A_i$.)  We maximize the
Likelihood function $L$ to obtain the \emph{Maximum Likelihood} values
of the parameters of the fitting function.  In practice we minimize
the negative of the log-likelihood, which is given by
\beq
 {\cal L} \equiv -\log(L) = \frac{1}{2} \sum_i \left ( \frac{ y_i -
 D_L^{\rm fit} (z_i;\,p_k) }{\sigma_{\rm dist}(z_i)} \right )^2 \,\,,
\eeq
where a constant term arising from the multiplicative factor is
ignored. We are interested in estimating the statefinder pair $r(z)$
and $s(z)$ and the deceleration parameter $q(z)$ from synthetic SNAP data. 

The priors that we have used for our reconstruction exercise are the
following.

The values of $H_0$ and $M_0$ (the absolute magnitude of SNe) are
assumed to be known. We consider a flat universe, so that the present
day value of $\omx $ is given by $\omx=1-\om=A_1+A_2+A_3$. Also, when
optimizing the model, we may assume priors on $\om$ using information
from other observations. This leaves only three free parameters
(including $\om$ on which bounds can be specified). (Optimizing
without priors we found the variances of $A_i$ to be much larger if no
bounds were specified on $\om$.)

{\bf Reconstruction of Cosmological Parameters}

Using the procedure described in detail above we now propose to
reconstruct different cosmologically important quantities using SNAP
data. We shall focus our attention to the statefinder pair $\lbrace
r(z), s(z)\rbrace$, the deceleration parameter $q(z)$ and the cosmic
equation of state $w(z)$.  Using SNAP specifications, we generated
$1000$ data sets $\{z_i^k,D_{Li}^k,\sigma_{zi}^k,\sigma_{D_{Li}}^k\}$,
where the index $k$ runs from $1$--$1000$ and the index $i$ from $1
\sim 2000$, with the LCDM as our fiducial model.  For each of these
experiments, the best-fit parameters, $A^j_k$ $(j=1,2)$, for the
polynomial fit to dark energy (\ref{eq:taylor}) were calculated. We
then calculated $r_k(z)$ and $s_k(z)$ for each experiment from the
calculated values $A^j_k$ .  The mean value of the statefinder pair
and other cosmological quantities is computed as
\ber\label{eq:rs_stats}
\langle r(z) \rangle &=& \frac{1}{1000}  \sum_{i=1}^{1000} r_i(z) \,\,,\nonumber\\
\langle s(z) \rangle &=& \frac{1}{1000} \sum_{i=1}^{1000} s_i(z) \,\,,
\eer
and so on for other quantities. Here the angular brackets denote
ensemble average. We may also calculate the covariance matrix of these
quantities at different redshifts which is given by
\beq
[ C_{ij} ] = \left( \begin{array}{cc} 
        C_{r r}  &      C_{r s} \\
        C_{r s}  &      C_{s s}
       \end{array} 
\right )\,\,,
\eeq
where 
\ber
C_{r r} &=&  \langle r(z)^2 \rangle - \langle r(z) \rangle^2 \,\,,\\
C_{s s} &=&  \langle s(z)^2 \rangle - \langle s(z) \rangle^2 \,\,,\\
C_{r s} &=&  \langle r(z) s(z) \rangle - \langle r(z) \rangle 
\langle s(z) \rangle \,\,,
\eer
and the angular averages are evaluates as in (\ref{eq:rs_stats}).

\begin{figure*}
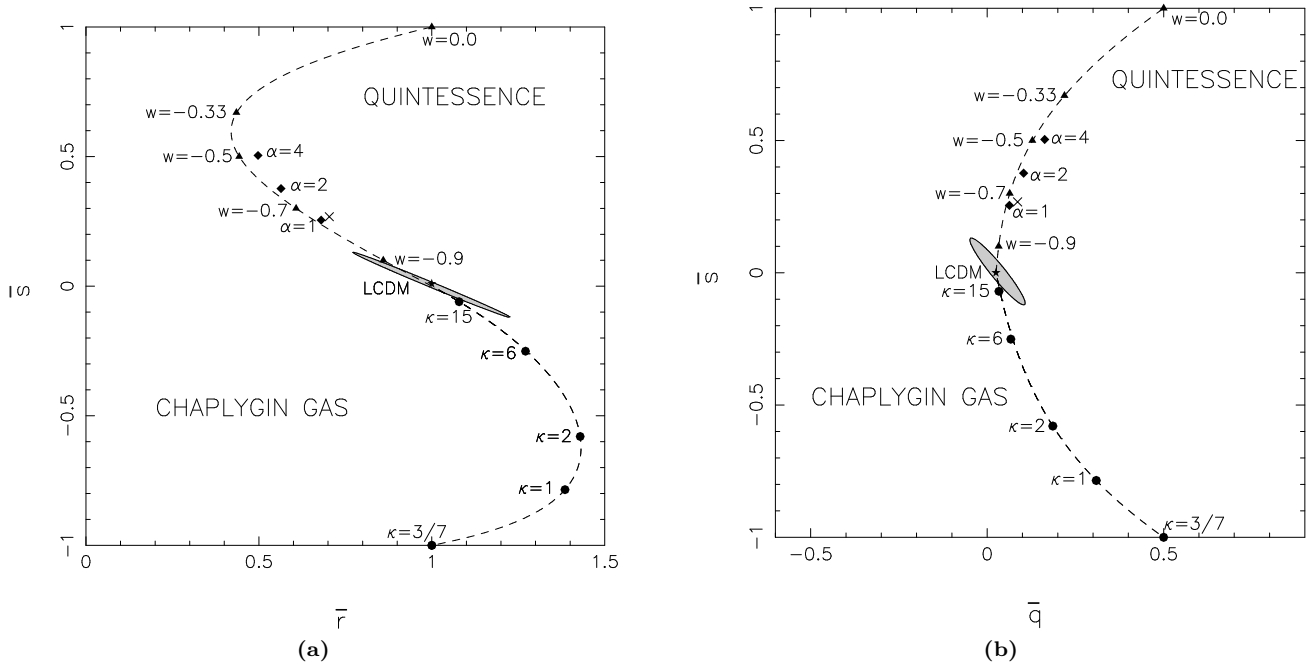
 
\centering
\begin{center}
$\begin{array}{c@{\hspace{0.4in}}c}
\multicolumn{1}{l}{\mbox{}} &
\multicolumn{1}{l}{\mbox{}} \\ [0.0cm]
\epsfxsize=3.2in
\epsffile{h_exp_rs_ex.epsi} &
\epsfxsize=3.2in
\epsffile{h_exp_qs_ex.epsi} \\  
\mbox{\bf (a)} & \mbox{\bf (b)}
\end{array}$
\end{center}
\caption{\small
This figure shows $3\sigma$ confidence levels in the averaged
statefinders (a) $\lbrace\sb,\rb\rbrace$ and (b)
$\lbrace\sb,\qb\rbrace$.  The polynomial fit to dark energy,
Eq~(\ref{eq:taylor}) has been used to reconstruct the statefinders for
an LCDM fiducial model with $\om = 0.3$.  The dashed line above the
LCDM fixed point represents the family of quiessence models having
$w=$ constant.  The dashed line below the LCDM fixed point shows
Chaplygin gas models.  It should be noted that the best-fit point in
both panels (a) \& (b) coincides with the LCDM fixed point (solid
star).  In the upper half of both panels, the solid rhombi correspond
to tracker potentials $V=V_0/\phi^{\alpha}$ while triangles show
$w$=constant quiessence models. In the lower half of both panels,
solid hexagons show Chaplygin gas models with different values of
$\kappa$. (The constant $\kappa$ gives the initial ratio between cold
dark matter and the Chaplygin gas. Only models with $\kappa \geq
\om/(1-\om)$ are permitted by theory, see Eq~(\ref{eq:chap_def}),
(\ref{eq:chap_lim}).)  All models, with the exception of the
braneworld model, have $\om=0.3$ currently. The braneworld model is
marked by a cross and corresponds to the DDG model
(\ref{eq:ddg}) with $\om=0.24$ which best-fits current
supernova data.  Comparing the left and right panels we find that
$\lbrace\sb,\qb\rbrace$ is a slightly better diagnostic than
$\lbrace\sb,\rb\rbrace$ for tracker and quiessence models and can be
used to rule out a constant equation of state $w\geq-0.9$ at the
$3\sigma$ level if the value of $\om$ is known exactly.  }
\label{fig:confid0}
\end{figure*}

\begin{figure*}
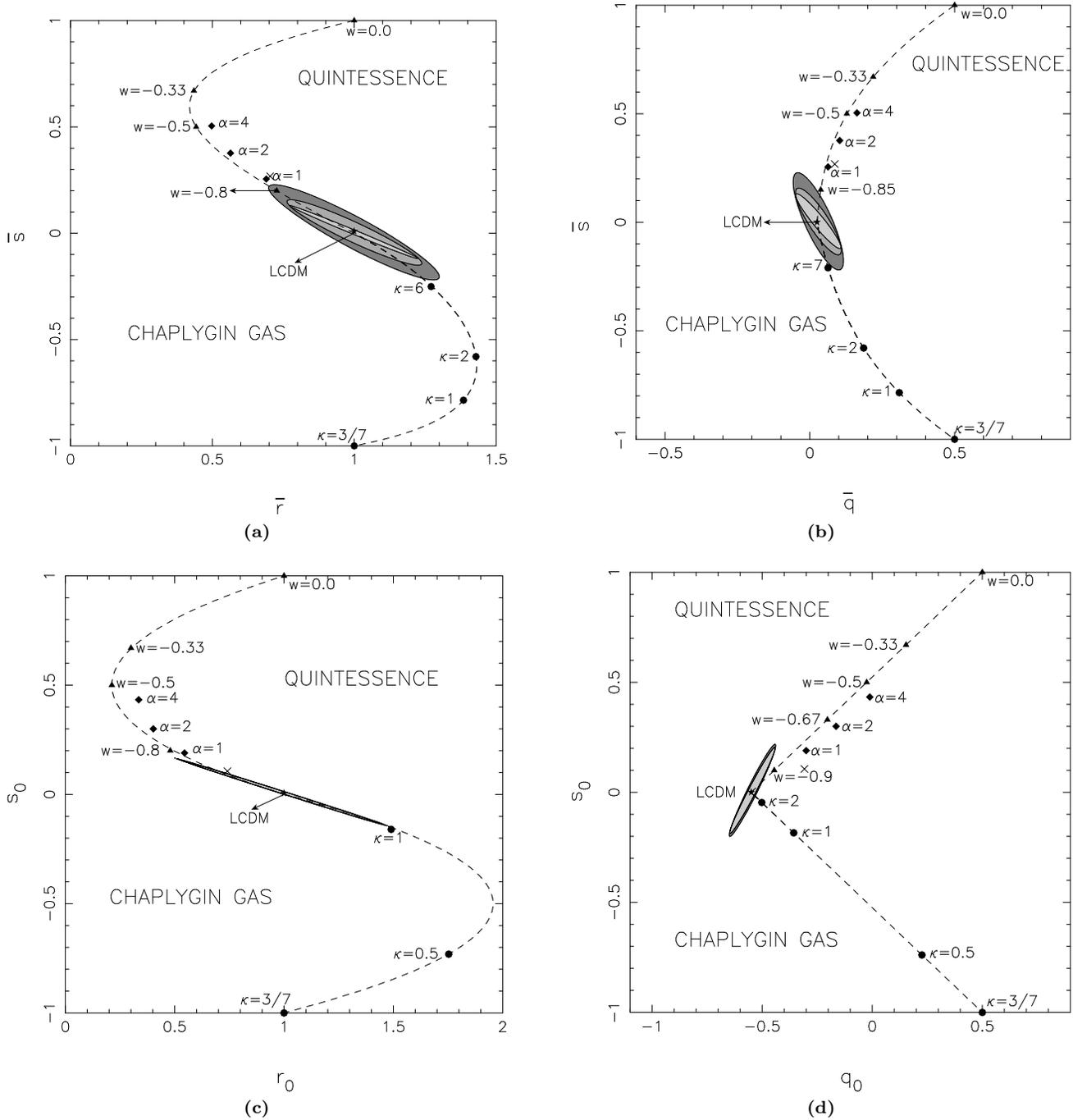
 
\centering
\begin{center}
$\begin{array}{c@{\hspace{0.4in}}c}
\multicolumn{1}{l}{\mbox{}} &
        \multicolumn{1}{l}{\mbox{}} \\ [0.0cm]
\epsfxsize=3.2in
\epsffile{h_exp_rs.epsi} &
        \epsfxsize=3.2in
        \epsffile{h_exp_qs.epsi} \\  
\mbox{\bf (a)} & \mbox{\bf (b)}
\end{array}$
$\begin{array}{c@{\hspace{0.4in}}c}
\multicolumn{1}{l}{\mbox{}} &
        \multicolumn{1}{l}{\mbox{}} \\ 
\epsfxsize=3.2in
\epsffile{h_exp_rs0.epsi} &
        \epsfxsize=3.2in
        \epsffile{h_exp_qs0.epsi} \\ 
\mbox{\bf (c)} & \mbox{\bf (d)}
\end{array}$
\end{center}
\caption{\small 
This figure shows $3\sigma$ confidence levels in the statefinders: (a)
$\lbrace \rb, \sb \rbrace$, (b) $\lbrace\qb,\sb\rbrace$, (c) $\lbrace
r_0, s_0\rbrace$ and (d) $\lbrace q_0, s_0\rbrace$.  The fiducial
model is assumed to be LCDM and, as in the previous figure, the
polynomial fit to dark energy, Eq~(\ref{eq:taylor}) is used to
reconstruct the statefinder pairs.  All notations are as in the
previous figure. {\em The current observational uncertainty in the
  value of the matter density is incorporated by marginalizing over
  the value of $\om$.}  The dark grey outer contour shows results for
the Gaussian prior $\om=0.3 \pm \sigma_{\om}$ with
$\sigma_{\om}=0.05$, the grey contour in the middle uses the Gaussian
prior $\sigma_{\om}=0.015$, and the light grey contour is when
$\om=0.3$ exactly.  Comparing panels (a) - (d) we find that $\lbrace
s_0, q_0\rbrace$ is an excellent diagnostic of quintessence models
which can be used to rule out a constant equation of state $w \geq
-0.9$ and tracker potentials $V(\phi) \propto \phi^{-\alpha}, ~\alpha
\geq 1$, at the $3\sigma$ level even if $\om$ is known to an accuracy
of only $\sim 17\%$.  It is important to note that of all statefinder
pairs $\lbrace s_0, q_0\rbrace$ is the least sensitive to the
uncertainty in the value of $\om$. This is reflected in the fact that
the $3\sigma$ confidence contour for $\lbrace s_0, q_0\rbrace$ with
$\om = 0.3 \pm 0.05$ is not very much larger than the $3\sigma$
confidence level obtained if $\om$ is known exactly ($\om = 0.3$).  On
the other hand the averaged statefinder pair $\lbrace\sb,\qb\rbrace$
is a very good diagnostic of Chaplygin gas models and rules out models
with $\kappa \leq 7$ at the $3\sigma$ level if $\om = 0.3 \pm 0.05$.
(The braneworld model marked by the cross can be ruled out by $\lbrace
s_0, q_0\rbrace$ as well as $\lbrace\sb,\qb\rbrace$.)  }
\label{fig:confid}
\end{figure*}

\section{Results and Discussion}
\label{sec:results}

From the results we can estimate the accuracy with which the ansatz
recovers model independent values of different cosmological
parameters, especially the statefinder pair introduced in
\citet{sssa02}.  We can also determine whether this pair is useful in
discriminating the cosmological constant model from other models of
dark energy.

\subsection{Cosmological reconstruction for an LCDM fiducial model} 
Synthetic supernova data are generated for a fiducial LCDM model with
$\Omega_\Lambda = 0.7, \om = 0.3$ and assuming SNAP specifications
summarized in the previous section.  Next we determine the statefinder
pair and other cosmological parameters as functions of the redshift
using the polynomial fit to dark energy (\ref{eq:taylor}).  Our
results can be represented in two complementary ways.  Firstly, we
show the confidence levels in the $r_0- s_0$ space, where the
subscript $'0'$ denotes the present day value of the statefinders.  We
also find it useful to consider the integrated, averaged quantities:
\ber
\qb &=& \frac{1}{z_{\rm max}}\int_0^{z_{\rm max}} q(z)\,dz\,\,,\\
\rb &=& \frac{1}{z_{\rm max}}\int_0^{z_{\rm max}} r(z)\,dz\,\,,\\
\sb &=& \frac{1}{z_{\rm max}}\int_0^{z_{\rm max}} s(z)\,dz\,\,.
\eer
For the LCDM model, $r$ and $s$ do not evolve with time, therefore we
find that $\rb=1$ and $\sb=0$. However for most other models of dark
energy the statefinder pair evolves and the averaged quantities differ
from their present day values. Due to averaging over redshift, the
averaged parameters $\rb,\sb$ are in many cases less noisy than
$r_0,s_0$. The maximum redshift used for our reconstruction is $z_{\rm
  max}=1.7$. One of the results of our analysis is that the
deceleration parameter $q$ is very well determined, see
Fig.~(\ref{fig:h_exp_q.ps}), therefore we also construct a second
statefinder pair, $\lbrace s,q \rbrace$, which will be shown to be an
excellent diagnostic of dark energy.

\begin{figure} 
\centering 
\centerline{ \psfig{figure=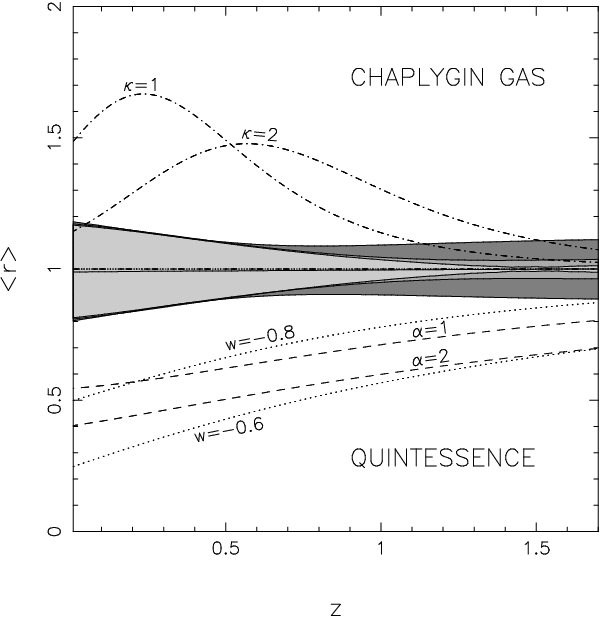,width=0.5\textwidth,angle=0} }
\caption{\small 
Variation of $<r(z)>$ with $z$ for the cosmological constant model.
Solid line shows best-fit $<r(z)>$ averaged over all realizations
calculated with the {\em polynomial fit to dark energy},
Eq~(\ref{eq:taylor}), for the prior $\om=0.3$ exactly. The
triple-dot-dashed line represents the exact value of $<r>=1$ for the
cosmological constant model. Shaded regions represent the $1 \sigma$
confidence levels for $<r(z)>$.  The dark grey outer contour is for
the Gaussian prior $\om=0.3 \pm \sigma_{\om}$ with
$\sigma_{\om}=0.05$, the grey contour in the middle uses the Gaussian
prior $\sigma_{\om}=0.015$, and the light grey contour uses $\om=0.3$
exactly. The dotted, dashed and dot-dashed lines represent the exact
values of $r(z)$ for different constant $w$ quiessence models, for
kinessence models with the tracker potential $V(\phi) = V_0/
\phi^{\alpha}$ for different values of $\alpha$, and for Chaplygin gas
models with different $\kappa$ respectively. We see that all the model
values plotted lie outside the $1 \sigma$ confidence level even for
the most conservative prior of $\sigma_{\om}=0.05$ at redshifts $\ggeq
0.3$.  }
\label{fig:h_exp_r.ps} 
\end{figure} 

\begin{figure} 
\centering 
\centerline{ \psfig{figure=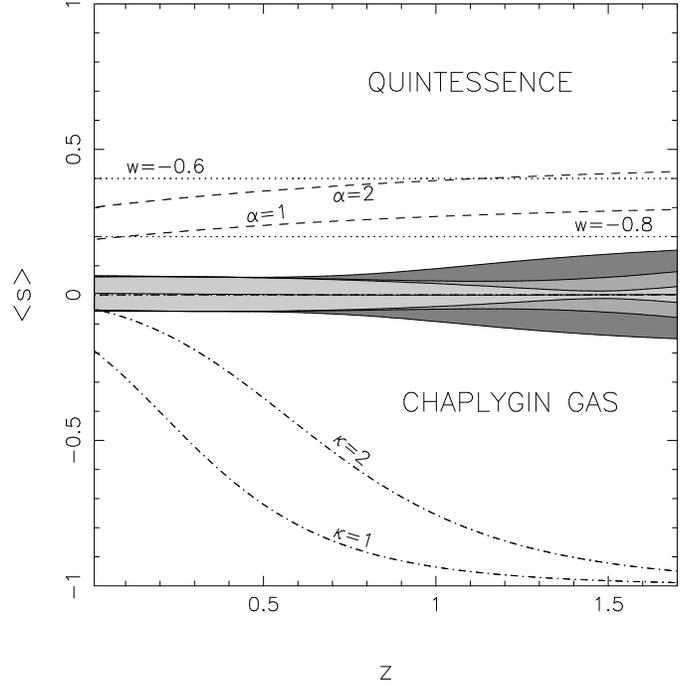,width=0.5\textwidth,angle=0} }
\caption{\small 
Variation of $<s(z)>$ with $z$ for the cosmological constant model.
Solid line shows best-fit $<s(z)>$ averaged over all realizations
calculated with the {\em polynomial fit to dark energy},
Eq~(\ref{eq:taylor}), for the prior $\om=0.3$ exactly. The
triple-dot-dashed line represents the exact value of $<s>=0$ for the
cosmological constant model. Shaded regions represent the $1 \sigma$
confidence levels for $<s(z)>$.  The dark grey outer contour is for
the Gaussian prior $\om=0.3 \pm \sigma_{\om}$ with
$\sigma_{\om}=0.05$, the grey contour in the middle uses the Gaussian
prior $\sigma_{\om}=0.015$, and the light grey contour uses $\om=0.3$
exactly. The dotted, dashed and dot-dashed lines represent the exact
values of $s(z)$ for different constant $w$ quiessence models, for
kinessence models with the tracker potential $V(\phi) = V_0/
\phi^{\alpha}$ for different values of $\alpha$, and for Chaplygin gas
models with different $\kappa$ respectively. We see that all the model
values plotted lie outside the $1 \sigma$ confidence level even for
the most conservative prior of $\sigma_{\om}=0.05$.}
\label{fig:h_exp_s.ps} 
\end{figure} 

Figure \ref{fig:confid0} shows the $99.73\%$ confidence level in
$\lbrace \sb,\rb \rbrace$ (left panel) and $\lbrace \sb, \qb \rbrace$
(right panel) for the fiducial LCDM model with $\om = 0.3$,
$\Omega_\Lambda = 0.7$. For comparison we also show values of $\rb,
\sb, \qb$ for quiessence, kinessence and Chaplygin gas models.  From
this figure we see that the statefinders can easily distinguish LCDM
from: (i) quiessence with $w \ggeq -0.9$ (ii) the Chaplygin gas with
$\kappa \lleq 15$ (iii) the quintessence potential $V(\phi) \propto
\phi^{-\alpha}, \alpha \ggeq 1$ and (iv) the DDG braneworld models
discussed in Deffayet \etal (2002).

The above analysis assumed that the value of $\om$ is known exactly.
However in practice it will be some time before $\om$ is known to
100\% accuracy and it is only natural to expect some amount of
uncertainty in the observational value of this important physical
parameter. We incorporate this uncertainty by marginalizing over the
value of $\om$. Two priors will be incorporated into our analysis, the
weak Gaussian prior: $\om = 0.3 \pm 0.05$ and the stronger Gaussian
prior: $\om = 0.3 \pm 0.015$.

Figures~ \ref{fig:confid} (a-d), show the confidence levels in the
statefinder pairs $\lbrace \sb, \rb \rbrace$, $\lbrace \sb, \qb
\rbrace$, $\lbrace s_0, r_0 \rbrace$ and $\lbrace s_0, q_0 \rbrace$
respectively. For purposes of discrimination we also show the values
of the $r, s, q$ for quiessence, kinessence, Chaplygin gas and
braneworld models. Figure~\ref{fig:confid}(a) shows that the
diagnostic $\lbrace \sb, \rb \rbrace$ permits the LCDM model to be
distinguished from quiessence models with $w \ggeq -0.8$, quintessence
models with $\alpha \ge 1$, Chaplygin gas models with $\kappa \lleq 6$
and braneworld models at the $99.73 \%$ confidence level and after
applying the strong Gaussian prior of $\om=0.3 \pm 0.015$.  The
discriminatory power of the statefinder clearly worsens for the weaker
prior $\om=0.3 \pm 0.05$.

The situation can be dramatically improved if, instead of working with
$\lbrace \sb, \rb \rbrace$ we use the diagnostic $\lbrace \sb, \qb
\rbrace$ (see figure \ref{fig:confid} (b)).  We find in this case that
the fiducial LCDM model can be distinguished from quiessence with $w
\ggeq -0.85$ and the braneworld model at the $99.73 \%$ confidence
level {\em even for} the weak prior $\om=0.3 \pm 0.05$.  In
Figures~\ref{fig:confid}(c) \& (d) we plot the confidence levels for
{\em current values} of the pair $\lbrace s_0, r_0 \rbrace$ and
$\lbrace s_0, q_0 \rbrace$. A few important points need to be noted
here:

\begin{figure*} \centering
\begin{center}
$\begin{array}{c@{\hspace{0.4in}}c}
\multicolumn{1}{l}{\mbox{}} &
\multicolumn{1}{l}{\mbox{}} \\ [0.0cm]
\epsfxsize=3.2in
\epsffile{h_exp_rs_comp1.epsi} &
\epsfxsize=3.2in
\epsffile{h_exp_rs_comp2.epsi} \\  
\mbox{\bf (a)} & \mbox{\bf (b)}
\end{array}$
\end{center}
\caption{\small 
$3\sigma$ confidence levels in the parameter space $\rb, \sb$ are
shown for the cosmological constant model and the $\alpha=1$ tracker
model using the {\em polynomial fit to dark energy},
Eq~(\ref{eq:taylor}). The solid stars represent the model value of the
parameter pair for the cosmological constant model and the $\alpha=1$
tracker kinessence models.  The dashed line above the LCDM point
represents the quiessence models, and that below the LCDM point
represents different Chaplygin gas models. The solid triangles
represent model values for constant $w$ quiessence models, and the
solid hexagons represent Chaplygin gas models with different values of
$\kappa$. Only those Chaplygin gas models with $\kappa \geq
\om/(1-\om)$ are allowed. For all the dark energy models, $\om=0.3$ is
used. The ellipses represent the $3 \sigma$ confidence levels in the
$\rb-\sb$ space for the exact prior $\om=0.3$. In (a), the dark grey
contour in solid outline represents the confidence level for the
cosmological constant fiducial model obtained when $r,s$ are averaged
over the redshift range $z=0$ to $z=1.7$.  The light grey contour with
dotted outline is the confidence level for the $\alpha=1$ tracker
kinessence model obtained when the averaging is over the entire
redshift range. In (b), we show confidence levels for the cosmological
constant model (dark grey contour) and for the $\alpha=1$ tracker
model (light grey contour) with the averaging done from $z=1$ to
$z=1.7$.  Remarkably, using the statefinders $\lbrace \rb, \sb
\rbrace$ one can rule out quintessence models with $w \ggeq -0.95$ and
Chaplygin gas models with $\kappa \lleq 24$ at $3\sigma$ if only very
high redshift SNe belonging to the redshift interval $1 \leq z \leq
1.7$ are considered. The reason for this is that both $r(z)$ and
$s(z)$ are determined to increasing accuracy at $z \ggeq 1$.  Indeed,
a `sweet spot' at $z_s \simeq 1.4$ ensures that both $r(z_s)$ \&
$s(z_s)$ are known with great accuracy at that point -- see figures
(~\ref{fig:h_exp_r.ps}) \& (~\ref{fig:h_exp_s.ps}).
}
\label{fig:sweet_rs} 
\end{figure*}

(i) the semi-major axis of the confidence ellipse for $\statek$ is
tilted away from the dashed curve representing constant $w$ models
(quiessence).  This enables the second statefinder pair $\statek$ to
be somewhat better at discriminating between LCDM and quintessence
models than $\statei$.  For instance, the current value $\lbrace s_0,
q_0 \rbrace$ can discriminate the cosmological constant model from
quiessence models having $w \ggeq -0.9$, kinessence models with
$\alpha \geq 1$, Chaplygin gas models with $\kappa \lleq 2$, and the
braneworld model {\em even after applying the weak Gaussian prior}
$\om = 0.3 \pm 0.05$.

(ii) For Chaplygin gas models averaging over redshift considerably
enhances the discriminatory prowess of both $\lbrace \sb, \rb \rbrace$
and $\lbrace \sb, \qb \rbrace$.

(iii) From figure (\ref{fig:confid}d) we find that marginalization
over $\om$ has only a small effect on the diagnostic $\lbrace s_0, q_0
\rbrace$ which contributes to making this statefinder pair a much
better discriminator of dark energy than $\lbrace r_0, s_0 \rbrace$ if
the value of $\om$ is uncertain.

Our results shown in figures~ \ref{fig:confid0} \& \ref{fig:confid}
clearly demonstrate that both $\statei$ as well as $\statek$ are
excellent diagnostics of dark energy with the latter being somewhat
more sensitive than the former.

We now proceed to examine the information content in the cosmological
parameters when examined individually.  In
Figure~\ref{fig:h_exp_r.ps}, we plot the variation of the ensemble
averaged value $\langle r(z) \rangle$ with redshift.  The $1\sigma$
error bounds are shown for two different priors on $\om$ and for the
case when the value of $\om$ is known exactly.  This figure shows that
$r(z)$ is a good diagnostic of dark energy and allows us to
discriminate (at the $68.3 \%$ CL) between different dark energy
models and LCDM. Discrimination improves at higher redshifts ($z \ggeq
0.8$) especially if the uncertainty in the value of $\om$ is small.
The sweet spot for this parameter, $\ie$ the point at which $r(z)$ is
most accurately determined, is at $z \sim 1.4$. \citep[For earlier
work on the sweet spot see][]{albrecht,turner,hut02}.

Figure~\ref{fig:h_exp_s.ps} shows the variation of the ensemble
averaged value of the second statefinder $\langle s(z) \rangle$ with
redshift. Again $1\sigma$ errors for the two priors $\om = 0.3 \pm
0.05$, $\om = 0.3 \pm 0.015$ and when the value of $\om$ is known
exactly ($\om = 0.3$) are shown.  We see that $s(z)$ is determined
even more accurately than $r(z)$, and therefore can serve as a better
diagnostic of dark energy.  For the strong Gaussian prior $\om = 0.3
\pm 0.015$, (or when $\om$ is known exactly) the value of $s$ is very
well determined even at higher redshifts, its sweet spot being at $z
\sim 1.4$.  For the weak prior $\om=0.3 \pm 0.05$, $s$ is not so well
determined at high redshifts, but it is still accurate enough to
distinguish between rival models of dark energy. Two points are of
interest here.  Firstly, $r$ and $s$ are both much more accurately
determined at higher redshifts if the value of $\om$ is accurately
known. This explains why the parameters $\lbrace \rb,\sb \rbrace$
perform better as discriminators than $\lbrace r_0,s_0 \rbrace$.
Secondly, the sweet spot for both these parameters appears at $z \sim
1.4$, {\em only if the value of $\om$ is accurately known}. Upon
marginalizing over $\om$ the sweet spot disappears both in the case of
$r(z)$ as well as in the case of $s(z)$.  Another point worth
mentioning is that Chaplygin gas models are much easier to rule out at
high $z$ than at low $z$, using either $r(z)$ or $s(z)$. As an
illustration, neither $r_0$ nor $s_0$ can distinguish a $\kappa = 2$
Chaplygin gas model from LCDM (with identical $\om$) at the $1\sigma$
level. However the {\em averaged-over-redshift} statefinders $\rb,
\sb$ can do so quite easily even at the $3 \sigma$ level, as
demonstrated in figures \ref{fig:confid0} and \ref{fig:confid}.

\begin{figure} 
\centering 
\centerline{ \psfig{figure=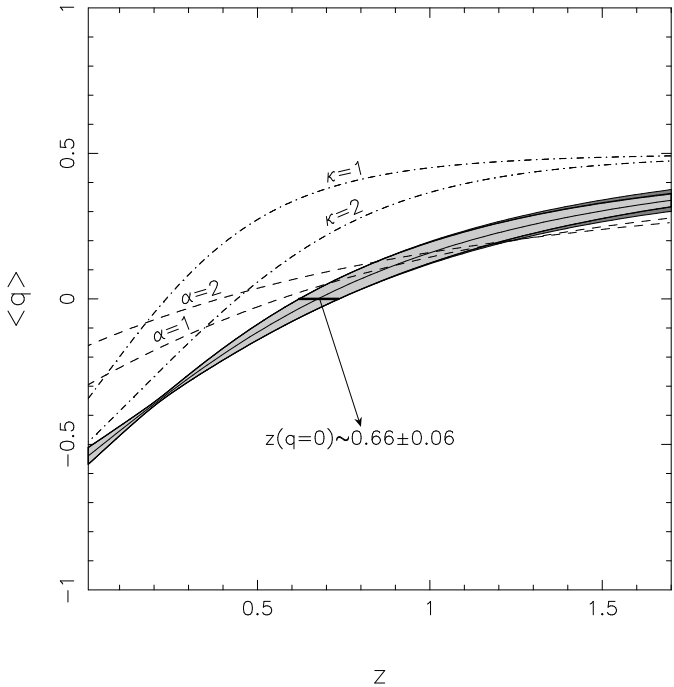,width=0.5\textwidth,angle=0} }
\caption{\small 
Variation of $<q(z)>$ with $z$ for the cosmological constant model.
Solid line shows best-fit $<q(z)>$ averaged over all realizations
calculated with the {\em polynomial fit to dark energy},
Eq~(\ref{eq:taylor}), for the prior $\om=0.3$ exactly. Shaded regions
represent the $1 \sigma$ confidence levels for $<q>$.  We find here
that the use of exact $\om=0.3$ and the Gaussian prior $\om=0.3 \pm
\sigma_{\om}$ with $\sigma_{\om}=0.015$ gives us almost the same
bounds, represented by the light grey contour, the dark grey outer
contour uses the Gaussian prior $\sigma_{\om}=0.05$. The dotted,
dashed and dot-dashed lines represent the model values of $<q>$ for
different constant $w$ quiessence models, for kinessence models with
the tracker potential $V(\phi) = V_0/ \phi^{\alpha}$ for different
values of $\alpha$, and for Chaplygin gas models with different
$\kappa$ respectively. The horizontal thick line represents the
accuracy with which the acceleration epoch is determined in this fit.
This figure demonstrates that $q$ can be used as a discriminator
between dark energy models at low redshifts$ \lleq 0.5$. Indeed the
location of a sweet spot at $z_s \simeq 0.2$ demonstrates that the
value of $q(z_s)$ is known to remarkably good accuracy !}
\label{fig:h_exp_q.ps} 
\end{figure} 

\begin{figure*} 
\centering 
\begin{center}
$\begin{array}{c@{\hspace{0.4in}}c}
\multicolumn{1}{l}{\mbox{}} &
\multicolumn{1}{l}{\mbox{}} \\ [0.0cm]
\epsfxsize=3.2in
\epsffile{h_exp_qs_comp1.epsi} &
\epsfxsize=3.2in
\epsffile{h_exp_qs_comp2.epsi} \\  
\mbox{\bf (a)} & \mbox{\bf (b)}
\end{array}$
\end{center}
\caption{\small 
$3\sigma$ confidence levels in the parameter space $\qb, \sb$ are
shown for the cosmological constant model and the $\alpha=1$ tracker
model using the {\em polynomial fit to dark energy},
Eq~(\ref{eq:taylor}). The solid stars represent the model value of the
parameter pair for the cosmological constant model and the $\alpha=1$
tracker kinessence models.  The dashed line above the LCDM point
represents the quiessence models, and that below the LCDM point
represents different Chaplygin gas models. The solid triangles
represent model values for constant $w$ quiessence models, and the
solid hexagons represent Chaplygin gas models with different values of
$\kappa$. Only those Chaplygin gas models with $\kappa \geq
\om/(1-\om)$ are allowed. For all the dark energy models, $\om=0.3$ is
used. The ellipses represent the $3 \sigma$ confidence levels in the
$\qb-\sb$ space for the exact prior $\om=0.3$. In (a), the dark grey
contour in solid outline represents the confidence level for the
cosmological constant fiducial model obtained when $q,s$ are averaged
in the redshift range $z=0$ to $z=1.7$.  The light grey contour with
dotted outline is the confidence level for the $\alpha=1$ tracker
kinessence model obtained when the averaging is over the entire
redshift range. In (b), we show confidence levels for the cosmological
constant model (dark grey contour) and for the $\alpha=1$ tracker
model (light grey contour) with the averaging done from $z=0$ to
$z=0.4$ for $q$ and from $z=1$ to $z=1.7$ for $s$. The reason for
choosing these ranges is that $q$ is extremely well-determined at low
redshifts, with a sweet spot at $z_s \simeq 0.25$, and $s$ is
accurately determined at high redshifts with a sweet spot at $z_s
\simeq 1.4$ -- see figures (~\ref{fig:h_exp_q.ps}) \&
(~\ref{fig:h_exp_s.ps}). Using the second statefinder pair $\lbrace
\qb, \sb \rbrace$ one can rule out quintessence models with $w \ggeq
-0.95$ and Chaplygin gas models with $\kappa \lleq 25$ at $3\sigma$ if
only very high redshift SNe belonging to the redshift interval $1 \leq
z \leq 1.7$ are considered for $s$ and low redshift SNe in the
interval $0 \leq z \leq 0.4$ are considered for $q$.
}
\label{fig:sweet_qs} 
\end{figure*}

Figure~\ref{fig:sweet_rs} shows how sweet spot information can be used
to improve the statefinders as a diagnostic tool.  For both $r$ and
$s$ the sweet spot appears at high redshifts. Therefore, one expects
that the discriminatory prowess of the statefinders will improve
considerably if only data at $z \geq 1$ is considered.  This is indeed
the case. Figure~\ref{fig:sweet_rs} shows $3 \sigma$ confidence levels
in $\lbrace \rb, \sb \rbrace$ for two cases: (a) the statefinder pair
is averaged over the full redshift range $0 \leq z \leq 1.7$, (b) the
statefinder pair is averaged over the high redshift range $1 \leq z
\leq 1.7$; $\om=0.3$ for both cases. The dark grey ellipses represent
the confidence level for the LCDM model, and the light grey ellipses
represent the confidence level for the $\alpha=1$ kinessence model. We
see that there is a dramatic improvement in the determination of the
statefinder pair in figure \ref{fig:sweet_rs}(b) where the
statefinders have been averaged only for $z \geq 1$.  From figure
\ref{fig:sweet_rs} (a) we see that $\lbrace \rb,\sb \rbrace$ can
discriminate between LCDM and quiessence models with $w \ggeq -0.90$
and Chaplygin gas models with $\kappa \lleq 15$, whereas
\ref{fig:sweet_rs}(b) shows that $\lbrace \rb,\sb \rbrace$ can
discriminate between LCDM and quiessence models with $w \ggeq -0.95$
and Chaplygin gas models with $\kappa \lleq 24$ !  We therefore
conclude that high redshift supernovae can play a crucial role in
constraining properties of dark energy and our results support the
views expressed in \citet{linder02}.  We must however note that in
order to use sweet spot information optimally the value of $\om$ must
be known to very high accuracy.  Indeed, for the Gaussian prior of
$\om=0.3 \pm 0.05$, a consideration of only high redshift supernovae
does not lead to any improvement in the results.  This is because, as
seen from Figures~\ref{fig:h_exp_r.ps} \& \ref{fig:h_exp_s.ps}, after
marginalization over $\om$, the sweet spot for both $r(z)$ and $s(z)$
disappears.  The second point to note is that the angle of inclination
of the semi-major axis of the ellipse to the $w=$ constant curve
(quiessence) appears to depend upon the redshift range over which the
statefinder pair is being averaged.

Figure~\ref{fig:h_exp_q.ps} shows the variation of the mean
deceleration parameter $q(z)$ with redshift. We see that $q(z)$ is
very well determined over the entire range $0 \leq z \leq 1.7$. This
justifies our choice of $\lbrace s,q \rbrace$ as the second
statefinder pair. Indeed, the behaviour of $r(z)$ and $s(z)$ on the
one hand and $q(z)$ on the other, is in some ways complementary. While
both $r(z)$ and $s(z)$ are determined to increasing accuracy at {\em
  higher redshifts}, the deceleration parameter is very well
determined at {\em lower redshifts} and the sweet spot for this
parameter appears at the redshift $z_s \simeq 0.25$. It is interesting
that, in sharp contrast to what was earlier observed for $r$ and $s$,
the sweet spot in $q(z)$ persists {\em even after we marginalize over
  $\om$} ! From figure~\ref{fig:h_exp_q.ps} we can also determine the
value of the acceleration epoch (the redshift at which the universe
began accelerating). We find that the acceleration epoch is determined
quite accurately: $z(q=0) = 0.66 \pm 0.06$.

\begin{figure} 
\centering 
\centerline{ \psfig{figure=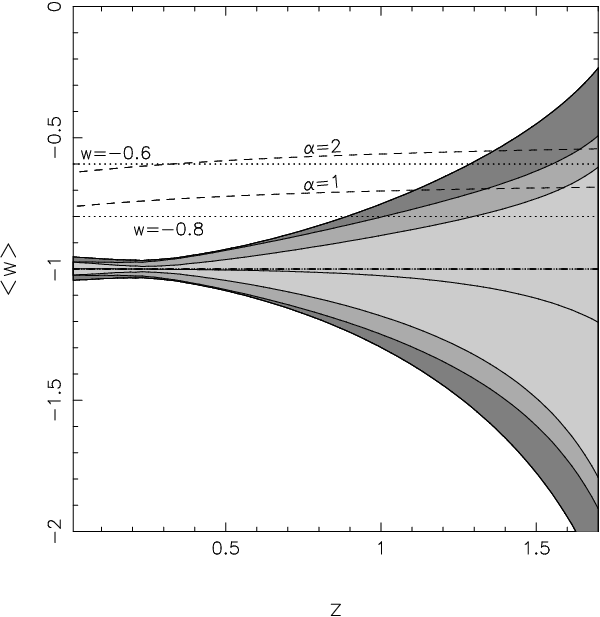,width=0.5\textwidth,angle=0} }
\caption{\small 
Variation of $<w(z)>$ with $z$ for the cosmological constant model.
Solid line shows best-fit $<w(z)>$ averaged over all realizations
calculated with the {\em polynomial fit to dark energy},
Eq~(\ref{eq:taylor}), for the prior $\om=0.3$ exactly. The dot-dashed
line represents the exact value of $<w>=-1$ for the cosmological
constant model. Shaded regions represent the $1 \sigma$ confidence
levels for $<w>$. The dark grey outer contour is for the Gaussian
prior $\om=0.3 \pm \sigma_{\om}$ with $\sigma_{\om}=0.05$, the grey
contour in the middle uses the Gaussian prior $\sigma_{\om}=0.015$,
and the light grey contour uses $\om=0.3$ exactly. The dotted and
dashed lines represent the model values of $<w>$ for different
constant $w$ quiessence models and for kinessence models with the
tracker potential $V(\phi) = V_0/ \phi^{\alpha}$ for different values
of $\alpha$ respectively. We see that $w$ can distinguish between the
cosmological constant model and other dark energy models only at $z
\lleq 1$. }
\label{fig:h_exp_w.ps} 
\end{figure} 

\begin{figure} 
\centering 
\centerline{ \psfig{figure=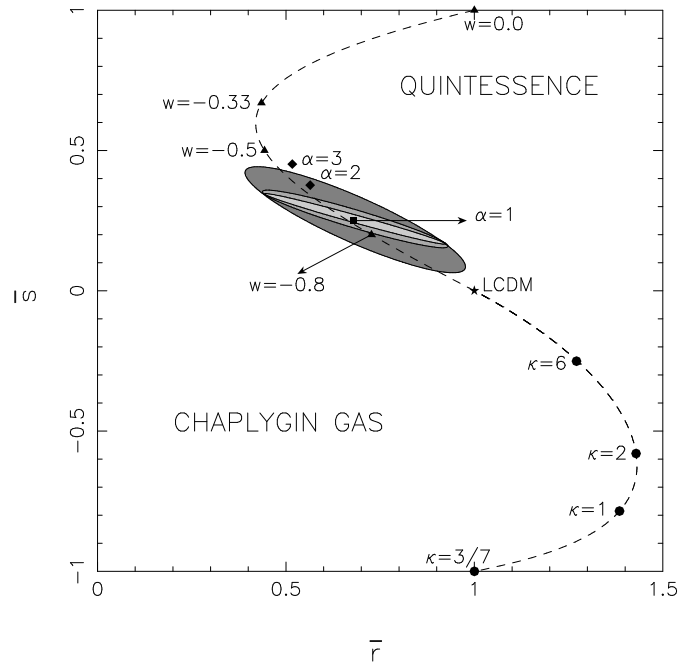,width=0.5\textwidth,angle=0} }
\caption{\small
$3\sigma$ confidence levels in the parameter space $\rb, \sb$ are shown
for the kinessence model with tracker potential
$V=V_0/\phi^{\alpha}$ for $\alpha=1$, for the {\em polynomial fit to
dark energy}, Eq~(\ref{eq:taylor}). The solid star represents the
model value of the parameter pair for the cosmological constant
model.  The dashed line above the LCDM point represents the
quiessence models, and that below the LCDM point represents
different Chaplygin gas models. The solid rhombi represent tracker
kinessence models with different $\alpha$, the solid triangles
represent the constant $w$ quiessence models, and the solid hexagons
represent Chaplygin gas models with different values of $\kappa$.
Only those Chaplygin gas models with $\kappa \geq \om/(1-\om)$ are
allowed. For all the dark energy models, $\om=0.3$ is used. The
square represents the best-fit for the $\alpha=1$ fiducial model.
The shaded ellipses represent the $3 \sigma$ confidence levels in
the $\rb-\sb$ space.  The dark grey outer contour is for the
Gaussian prior with $\sigma_{\om}=0.05$, the grey contour in the
middle uses the Gaussian prior $\sigma_{\om}=0.015$, and the light
grey contour uses $\om=0.3$ exactly. }
\label{fig:h_exp_rs1.ps}
\end{figure}

\begin{figure} 
\centering 
\centerline{ \psfig{figure=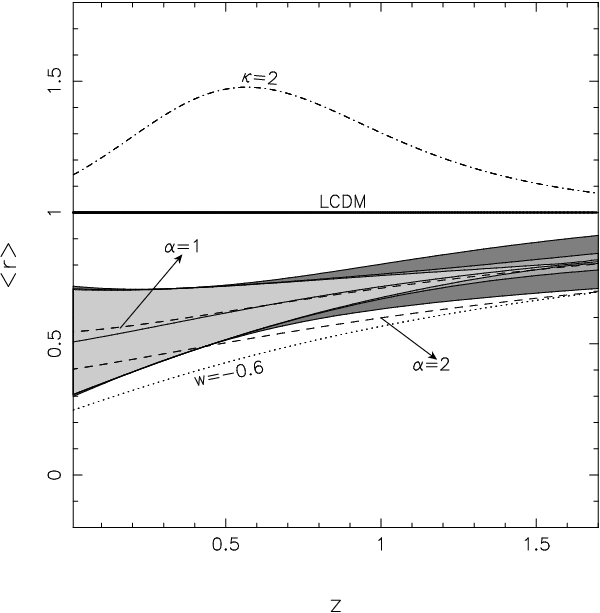,width=0.5\textwidth,angle=0} }
\caption{\small
Variation of $<r(z)>$ with $z$ for the kinessence model with the
tracker potential $V(\phi) = V_0/ \phi^{\alpha}$ for $\alpha=1$.
Solid line shows best-fit $<r(z)>$ averaged over all realizations
calculated with the {\em polynomial fit to dark energy},
Eq~(\ref{eq:taylor}), for the prior $\om=0.3$ exactly. Shaded regions
represent the $1 \sigma$ confidence levels for $<r>$.  The dark grey
outer contour is for the Gaussian prior $\om=0.3 \pm \sigma_{\om}$
with $\sigma_{\om}=0.05$, the grey contour in the middle uses the
Gaussian prior $\sigma_{\om}=0.015$, and the light grey contour uses
$\om=0.3$ exactly. The dotted, dashed and dot-dashed lines represent
the model values of $<r>$ for different constant $w$ quiessence
models, for tracker kinessence models for different values of
$\alpha$, and for Chaplygin gas models with different $\kappa$
respectively. The thick solid line represents LCDM. }
\label{fig:h_exp_r1.ps}
\end{figure}

\begin{figure} 
\centering 
\centerline{ \psfig{figure=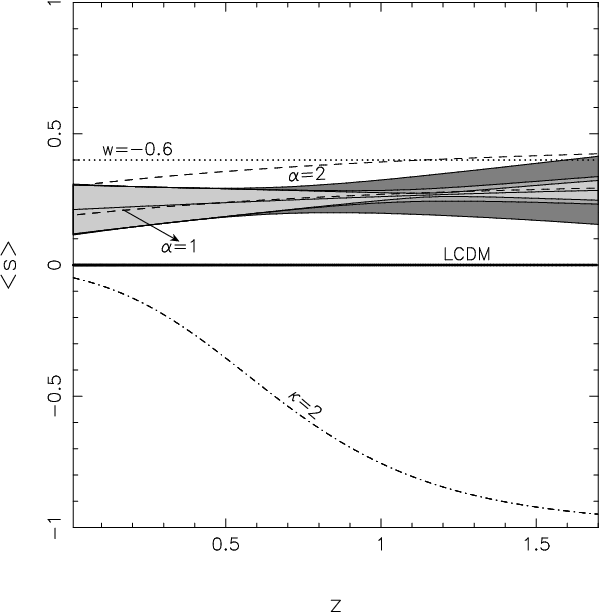,width=0.5\textwidth,angle=0} }
\caption{\small
Variation of $<s(z)>$ with $z$ for the kinessence model with the
tracker potential $V(\phi) = V_0/ \phi^{\alpha}$ for $\alpha=1$.
Solid line shows best-fit $<s(z)>$ averaged over all realizations
calculated with the {\em polynomial fit to dark energy},
Eq~(\ref{eq:taylor}), for the prior $\om=0.3$ exactly. Shaded regions
represent the $1 \sigma$ confidence levels for $<s>$.  The dark grey
outer contour is for the Gaussian prior $\om=0.3 \pm \sigma_{\om}$
with $\sigma_{\om}=0.05$, the grey contour in the middle uses the
Gaussian prior $\sigma_{\om}=0.015$, and the light grey contour uses
$\om=0.3$ exactly. The dotted, dashed and dot-dashed lines represent
the model values of $<s>$ for different constant $w$ quiessence
models, for tracker kinessence models for different values of
$\alpha$, and for Chaplygin gas models with different $\kappa$
respectively. The thick solid line represents LCDM. }
\label{fig:h_exp_s1.ps}
\end{figure}

\begin{figure} 
\centering 
\centerline{ \psfig{figure=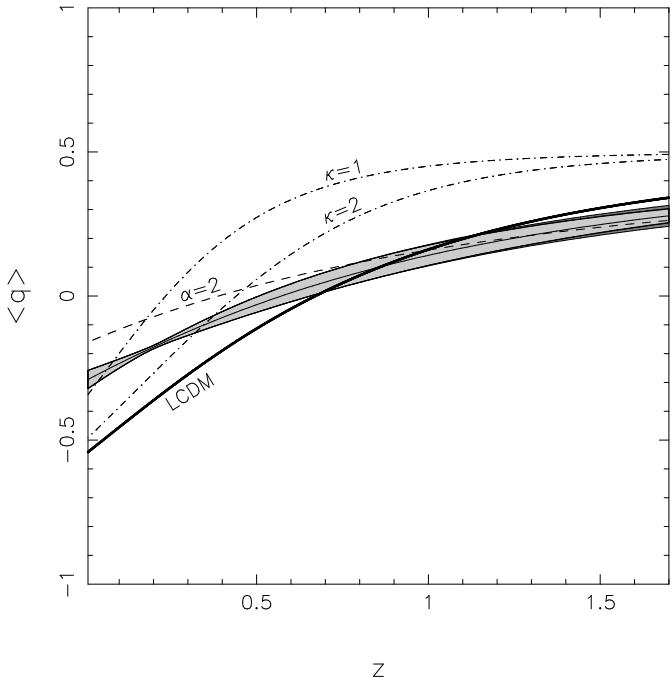,width=0.5\textwidth,angle=0} }
\caption{\small
Variation of $<q(z)>$ with $z$ for the kinessence model with the
tracker potential $V(\phi) = V_0/ \phi^{\alpha}$ for $\alpha=1$.
Solid line shows best-fit $<q(z)>$ averaged over all realizations
calculated with the {\em polynomial fit to dark energy},
Eq~(\ref{eq:taylor}), for the prior $\om=0.3$ exactly. Shaded regions
represent the $1 \sigma$ confidence levels for $<q>$. We find here
that the use of exact $\om=0.3$ and the Gaussian prior $\om=0.3 \pm
\sigma_{\om}$ with $\sigma_{\om}=0.015$ gives us almost the same
bounds, represented by the light grey contour, the dark grey outer
contour uses the Gaussian prior $\sigma_{\om}=0.05$. The dashed and
dot-dashed lines represent the model values of $<s>$ for tracker
kinessence models for different values of $\alpha$, and for Chaplygin
gas models with different $\kappa$ respectively. The thick solid line
represents LCDM. }
\label{fig:h_exp_q1.ps}
\end{figure}

\begin{figure} 
\centering 
\centerline{ \psfig{figure=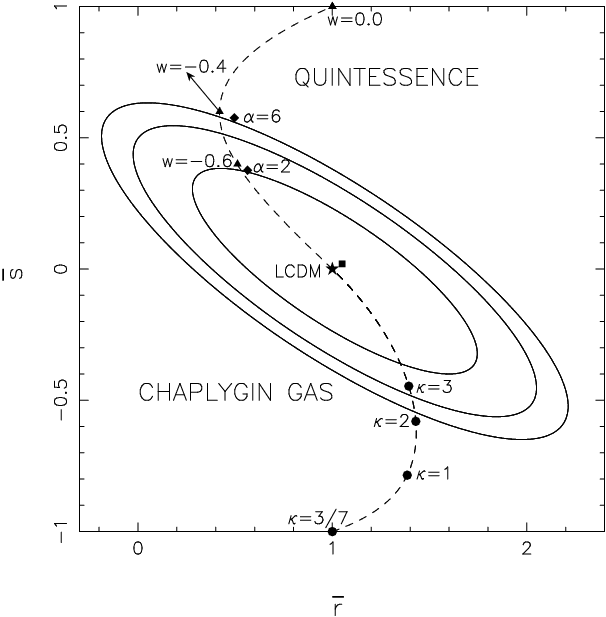,width=0.5\textwidth,angle=0} }
\caption{\small 
$3\sigma$ confidence levels in the parameter space $\rb, \sb$ are
shown for the cosmological constant model using the {\em polynomial
  fit to luminosity distance}, Eq~(\ref{eq:poly}). The solid star
represents the model value of the parameter pair for the cosmological
constant model.  The dashed line above the LCDM point represents the
quiessence models, and that below the LCDM point represents different
Chaplygin gas models. The solid rhombi represent tracker kinessence
models with different $\alpha$, the solid triangles represent model
values for constant $w$ quiessence models, and the solid hexagons
represent Chaplygin gas models with different values of $\kappa$. Only
those Chaplygin gas models with $\kappa \geq \om/(1-\om)$ are allowed.
For all the dark energy models, $\om=0.3$ is used. The best-fit point
for the reconstruction is represented by the solid square. The
ellipses represent the $1 \sigma$, $2 \sigma$, $3 \sigma$ confidence
levels in the $\rb-\sb$ space. }
\label{fig:dl_exp_rs.ps} 
\end{figure}

\begin{figure} 
\centering 
\centerline{ \psfig{figure=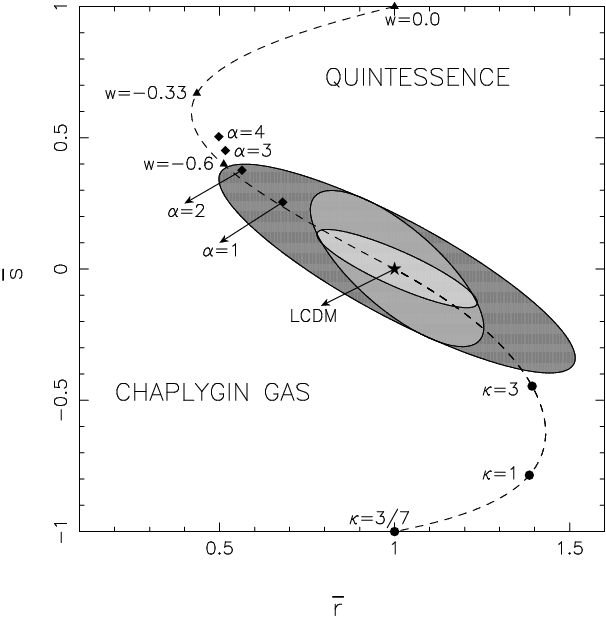,width=0.5\textwidth,angle=0} }
\caption{\small 
$3\sigma$ confidence levels in the parameter space $\rb, \sb$ are
shown for the cosmological constant model using the {\em polynomial
  fit to equation of state}, Eq~(\ref{eq:wpol}). The solid star
represents the model value of the parameter pair for the cosmological
constant model.  The dashed line above the LCDM point represents the
quiessence models, and that below the LCDM point represents different
Chaplygin gas models. The solid rhombi represent tracker kinessence
models with different $\alpha$, the solid triangles represent model
values for constant $w$ quiessence models, and the solid hexagons
represent Chaplygin gas models with different values of $\kappa$. Only
those Chaplygin gas models with $\kappa \geq \om/(1-\om)$ are allowed.
For all the dark energy models, $\om=0.3$ is used. The best-fit point
lies on the LCDM point for this reconstruction. The shaded ellipses
represent the $3 \sigma$ confidence levels in the $\rb-\sb$ space.
The dark grey outer contour is for the Gaussian prior $\om=0.3 \pm
\sigma_{\om}$ with $\sigma_{\om}=0.05$, the grey contour in the middle
uses the Gaussian prior $\sigma_{\om}=0.015$, and the light grey
contour uses $\om=0.3$ exactly. }
\label{fig:w_exp_rs.ps} 
\end{figure}

Figure~\ref{fig:sweet_qs} shows maximum likelihood contours for the
pair $\lbrace \sb,\qb \rbrace$ where $\qb$ has been averaged over the
redshift interval $0 < z \leq 0.4$ while $\sb$ has been averaged over
$1 \leq z \leq 1.7$.  This figure clearly demonstrates that $\lbrace
\sb,\qb \rbrace$ is an excellent diagnostic of dark energy since it
can distinguish LCDM from quiessence models with $w \geq -0.95$ on the
one hand and from Chaplygin gas models with $\kappa \leq 25$ on the
other (both at the $99.73 \%$ CL).  Figures~\ref{fig:sweet_rs}
and~\ref{fig:sweet_qs} show that the ability of the averaged
statefinder pairs $\lbrace \sb, \rb \rbrace$ and $\lbrace \sb, \qb
\rbrace$ to discriminate between dark energy models is comparable if
the value of $\om$ is known exactly.  (As demonstrated earlier in
figure (\ref{fig:confid}), $\statek$ is a more sensitive diagnostic
than $\statei$ if we marginalize over $\om$.)

Figure~\ref{fig:h_exp_w.ps} shows how the equation of state $w(z)$
varies with redshift. We see that, although the equation of state is
determined remarkably well at small redshifts, cosmological
reconstruction of $w(z)$ steadily worsens with $z$ and deteriorates
rapidly beyond $z \simeq 1$.  With the most conservative prior $\om =
0.3 \pm 0.05$, the ansatz (\ref{eq:taylor}) can distinguish between
the cosmological constant model and the quiessence model with $w =
-0.96$ at the $3\sigma$ level provided we restrict ourselves to low
redshifts $z \lleq 0.4$.  For higher redshifts, the LCDM model cannot
be distinguished from the $w = -0.8$ model beyond $z \simeq 1.1$
(after marginalizing over $\om$ with the prior $\om = 0.3 \pm 0.05$).
In the ideal case when $\om=0.3$ exactly, LCDM and the $w = -0.8$
model can be distinguished up to $z \sim 1.3$ but not beyond.  (In
this case the best-fit $w(z)$ is biased beyond $z \simeq 0.7$ and
takes on a lower value than the fiducial $w=-1$.)  Somewhat
surprisingly, although $w(z)$ and $q(z)$ are related through
(\ref{eq:state}) and therefore carry essentially the same information,
even a cursory examination of figures \ref{fig:h_exp_q.ps} and
\ref{fig:h_exp_w.ps} reveals that the ansatz (\ref{eq:taylor}) does
not reconstruct $w(z)$ to the same accuracy as it reconstructs $q(z)$.
However, like $q(z)$, $w(z)$ is reasonably well determined at low
redshifts, having a sweet spot at $z \sim 0.25$. The sweet spot
persists when we marginalize over $\om$ using the Gaussian prior
$\om=0.3 \pm 0.015$ but disappears when the uncertainty in $\om$ is
increased using the prior $\om=0.3 \pm 0.05$.

\subsection{Cosmological reconstruction for a tracker model} 
We now briefly examine the accuracy of the statefinder pair and the
ansatz (\ref{eq:taylor}) in determining the statefinder pair for a
fiducial model other than LCDM.  We know that the ansatz returns exact
values for the cosmological constant $w=-1$ as well as for quiessence
having the constant equation of state $w=-2/3$ and $w=-1/3$ (see
figure \ref{fig:h_exp_err.ps}).  It is therefore important to study
the accuracy of the statefinder in reconstructing the properties of
dark energy in models in which both the dark energy density as well as
the equation of state vary with time and for which the ansatz
(\ref{eq:taylor}) is approximate.  For this purpose we shall work with
a fiducial dark energy model which evolves under the influence of the
tracker potential $V=V_0/\phi$ and use the ansatz (\ref{eq:taylor}) in
tandem with the statefinders (\ref{eq:statefinder}) to reconstruct the
properties of dark energy.

Figure~\ref{fig:h_exp_rs1.ps} shows our results in terms of $3\sigma$
confidence levels in $\lbrace \rb, \sb \rbrace$. We find that, if the
value of $\om$ is known to reasonable accuracy ($\om=0.3 \pm 0.015$)
then the averaged statefinder pair $\lbrace \rb, \sb \rbrace$ is able
to distinguish the model $V=V_0/\phi$ from the model $V=V_0/\phi^2$ at
the $3\sigma$ level.  As expected, a large uncertainty in the current
value of the matter density reduces the efficiency of this diagnostic
and the two models $V=V_0/\phi$ and $V=V_0/\phi^2$ cannot be clearly
distinguished if the uncertainty in $\om$ is increased to $\om=0.3 \pm
0.05$.

Figures~\ref{fig:h_exp_r1.ps} \& \ref{fig:h_exp_s1.ps} show the
performance of the ensemble averaged statefinders $\langle
r(z)\rangle$ and $\langle s(z) \rangle$. As in the earlier case when
our fiducial model was assumed to be LCDM, we find that the errors in
$r(z)$ and $s(z)$ are small.  However a slight bias in the
determination of the statefinders exists at low redshifts so that for
$z \lleq 0.4$ the value of the best fit $\langle r \rangle$ ($\langle
s \rangle$) is larger (smaller) than the exact fiducial value.
Averaging over the entire redshift range can significantly reduce this
bias and we conclude that the ansatz (\ref{eq:taylor}) works well even
for those dark energy models for which it does not return exact
values.  One should also note the reappearance of the sweet spot for
the statefinders $r$ \& $s$ in the figures~\ref{fig:h_exp_r1.ps} \&
\ref{fig:h_exp_s1.ps}.  For the statefinder $r(z)$ the sweet spot
appears at $z \simeq 1.6$, while for $s(z)$ the sweet spot is at $z
\simeq 1.2$.  As in the case of the LCDM model, one can try and
constrain the properties of dark energy further by evaluating the
averaged statefinders using SNe data only from $z \geq 1$. Our results
shown in figure~\ref{fig:sweet_rs} demonstrate that the confidence
ellipse for $\lbrace \sb, \rb \rbrace$ becomes much smaller when the
averaging is done over the redshift range $1 \leq z \leq 1.7$ than
when the averaging is over the entire redshift range.

The performance of the deceleration parameter $\langle q(z) \rangle$
for this quintessence model is shown in the
figure~\ref{fig:h_exp_q1.ps}. We see that the deceleration parameter
is very accurately determined.  The sweet spot for the deceleration
parameter occurs at lower redshifts, at $z \simeq 0.2$, and by
combining higher redshift data in determining $\sb$ with lower
redshift data for determining $\qb$ we can significantly improve the
errors on the second statefinder pair $\statek$, as demonstrated in
figure~\ref{fig:sweet_qs}. As was the case for the LCDM model, the
sweet spot gradually disappears if we incorporate the prevailing
uncertainty in the value of the matter density by marginalizing over
large values of $\om$.

\subsection{Cosmological reconstruction using other fitting functions} 
For comparison, we also carry out the reconstruction exercise using
two of the fitting functions described in section
\ref{sec:cosmo_fits}.  In figure~\ref{fig:dl_exp_rs.ps} we show the
results for $\lbrace \rb,\sb \rbrace$ using the polynomial fit to the
luminosity distance (\ref{eq:poly}) with $N=5$. In this case, because
of the nature of the ansatz, it is not possible to place any priors on
$\om$.  We find that this ansatz does not perform well for the
statefinder pair. Firstly, it does not determine $\rb,\sb$ with the
accuracy seen in the case of the polynomial fit to dark energy.
Secondly, the best-fit value for $\lbrace \rb,\sb \rbrace$ is biased
with respect to the fiducial LCDM value.  Additionally, the errors on
both $r$ and $s$ are unacceptably large due to which one cannot
distinguish between the cosmological constant model and kinessence
models with $\alpha < 6$, quiessence models with $w \lleq -0.4$, and
Chaplygin gas models with $\kappa \ggeq 2$ at the $3 \sigma$
confidence level.  Even at $1 \sigma$ ($68.3 \% $ CL), one can only
discriminate LCDM from kinessence models with $\alpha \geq 3$,
quiessence models with $w \ggeq -0.6$, and Chaplygin gas models with
$\kappa \lleq 3$.  We therefore conclude that the polynomial fit to
the luminosity distance (\ref{eq:poly}) is not very useful for the
reconstruction of the statefinders.

We also carry out a similar reconstruction exercise using the
polynomial fit to the equation of state (\ref{eq:wpol}) with $N=1$.
This ansatz can accommodate priors on $\om$ and we expect it to
perform better than the polynomial fit to the luminosity distance.
Indeed, figure~\ref{fig:w_exp_rs.ps} clearly demonstrates that a two
parameter Taylor expansion in the equation of state is better than a
five parameter expansion in the luminosity distance \citep[Our
results in this context support the earlier observations
of][]{albrecht}. From figure \ref{fig:w_exp_rs.ps} we find that the
ansatz (\ref{eq:wpol}) can distinguish the cosmological constant model
from quiessence models with $w \ggeq -0.6$, kinessence models with
$\alpha \geq 3$, and Chaplygin gas models with $\kappa \lleq 3$ at the
$99.73 \%$ confidence limit after we have marginalized over $\om$ with
a Gaussian prior of $\om=0.3 \pm 0.05$. However a comparison of figure
\ref{fig:w_exp_rs.ps} with figure \ref{fig:confid} shows that the
equation of state expansion (\ref{eq:wpol}) is not quite as accurate
as the polynomial fit to dark energy (\ref{eq:taylor}) in
reconstructing the statefinder pair $\lbrace \rb, \sb \rbrace$.
We therefore conclude that the statefinders can be reconstructed using
several complementary methods.  The polynomial fit for dark energy
(\ref{eq:taylor}), by providing a good reconstruction of the
parameters $\rb, \sb, \qb$, can successfully be used for the model
independent reconstruction of dark energy.

\section{\bf Conclusions and discussion} 

This paper contains an in depth study of properties of the statefinder
diagnostic introduced in \citet{sssa02}. The statefinder pairs
$\statei$ and $\statek$ have the potential to successfully
discriminate between a wide variety of dark energy models including
the cosmological constant, quintessence, the Chaplygin gas and
braneworld models.  The statefinders play a particularly important
role in the case of modified gravity theories such as string/M-theory
inspired scalar-tensor models and braneworld models of dark energy,
for which the equation of state is not a fundamental physical entity
and therefore does not provide us with an adequate description of the
accelerating universe.  Our results, summarized in figures
\ref{fig:evolve} and \ref{fig:brane}, show that the statefinders $r$,
$s$ considerably extend and supplement traditional measures of
cosmological dynamics such as the deceleration parameter $q$.  To give
a concrete example of how this can happen consider two (or more)
cosmological dark energy models which have identical (hence
degenerate) values of $q_0$. Although such models will have the same
current value of ${\ddot a}/a$, the value of the third derivative
$\atridot$ (hence $r$ \& $s$) will in general be different in both
models.  The statefinder pairs $\statej$ and $\statek$ therefore
provide us with a `phase-space' picture of dark energy which
distinguishes dynamical dark energy models both from each other as
well as from the cosmological constant and helps break cosmological
degeneracies present in rival models of dark energy.  The statefinder
$s$ is remarkably sensitive to the {\em total} pressure of {\em all}
forms of matter and radiation in the universe.  As a result $s$
remains sensitive to the presence of dark energy even at moderately
high redshifts $z \sim 10$, when the universe is matter dominated.

Forthcoming space-based missions such as SNAP are expected to greatly
increase and improve the current Type Ia supernova inventory.
Anticipating this development we have carried out a maximum likelihood
analysis which combines the statefinder diagnostic with realistic
expectations from the SNAP experiment.  Our results, summarized in
figures \ref{fig:confid0} - \ref{fig:sweet_qs}, show that both
$\statei$ as well as $\statek$ are excellent diagnostics of dark
energy.  If the value of $\om$ is known exactly, then the
averaged-over-redshift statefinder pair $\lbrace \sb, \qb, \rbrace$
can distinguish between the cosmological constant model ($w = -1$) and
a dark energy model having $w = -0.9$ at the 99.73\% CL. It can also
distinguish (at the same level of confidence) the cosmological
constant (LCDM) from Chaplygin gas models with $\kappa \leq 7$.

Keeping in mind the fact that the current observational data determine
$\om$ to a finite level of accuracy, we have probed how well the
statefinder fares as a diagnostic after one incorporates the
prevailing uncertainty in the value of the matter density by
marginalizing over values of $\om$ which are uncertain.  Somewhat
surprisingly, the statefinder fares rather well even for the
relatively weak prior $\om = 0.3 \pm 0.05$. In this case, by employing
the diagnostic $\lbrace s_0, q_0 \rbrace$, the LCDM model can be
differentiated from the $w= -0.9$ model on the one hand, and from the
tracker potential $V(\phi) \propto \phi^{-1}$ and the DDG braneworld
model (Deffayet \etal 2002) on the other, at the 99.73\% CL.

Finally we should mention that of the two statefinders $s(z)$ appears
to be better constrained by observations especially if the uncertainty
in $\om$ is small. Interestingly both $r(z)$ and $s(z)$ show less
scatter at higher redshifts ($z \ggeq 1$) and thereby complement the
behaviour of the deceleration parameter $q(z)$ and the cosmic equation
of state $w(z)$ which are better constrained at lower $z$ ($z \lleq
0.4$). One is therefore tempted to construct a new diagnostic $\lbrace
\sb, \qb \rbrace$, where $\sb$ is averaged over the redshift range $ 1
\leq z \leq 1.7$ whereas $\qb$ is averaged over the redshift range $ 0
< z \leq 0.4$.  From figures \ref{fig:sweet_rs} \& \ref{fig:sweet_qs}
we find that $\lbrace \sb, \qb \rbrace$ provides an extremely potent
diagnostic of dark energy since it can distinguish a fiducial LCDM
model from a dark energy model with $w \geq -0.95$ on the one hand,
and from a Chaplygin gas model with $\kappa \leq 25$ on the other, at
the 99.73\% confidence level.

We therefore believe we have convincingly demonstrated that the
statefinder pair $\lbrace \sb, \rb \rbrace$ and $\lbrace \sb, \qb
\rbrace$ provide an excellent diagnostic of dark energy which will be
used to successfully differentiate between the cosmological constant
and dynamical models of dark energy.

\[ \]
\medskip
\noindent{\it Acknowledgments:}

We benefited from useful discussions with Dima Pogosyan.  VS and AS
acknowledge support from the ILTP program of cooperation between India
and Russia. AS acknowledges the hospitality of IUCAA where this work
was completed.  UA thanks the CSIR for providing support for this work.
AS was partially supported by the Russian Foundation for Basic
Research, grant 02-02-16817, and by the Research Program "Astronomy"
of the Russian Academy of Sciences.

\end{document}